\documentstyle[twocolumn,psfig]{mn}
\def\spose#1{\hbox to 0pt{#1\hss}}
\def\lta{\mathrel{\spose{\lower 3pt\hbox{$\sim$}}
    \raise 2.0pt\hbox{$<$}}}
\def\gta{\mathrel{\spose{\lower 3pt\hbox{$\sim$}}
    \raise 2.0pt\hbox{$>$}}}
\def\arcsec{\hbox{$^{\prime\prime}$}}
\def\farcm{\hbox{$.\mkern-4mu^\prime$}}
\def\farcs{\hbox{$.\!\!^{\prime\prime}$}}
\def\kms{$\mbox{km s}^{-1}$}

\def\sauron{{\tt SAURON}}
\def\oasis{{\tt OASIS}}
\def\tiger{{\tt TIGER}}

\def\isis{{\tt ISIS}}
\def\fos{{\tt FOS}}
\def\stis{{\tt STIS}}
\title[The SAURON project. II.] {The SAURON project. II. Sample
                                 and early results}
\author[P.T.\ de Zeeuw et al.]  {P.\ Tim de Zeeuw$^1$\thanks{E-mail:
       tim@strw.leidenuniv.nl}, M.\ Bureau$^1$, Eric Emsellem$^2$,
       R.\ Bacon$^2$,
\newauthor C.\ Marcella Carollo$^3$, Y.\ Copin$^{1, 2}$, Roger L.\ Davies$^4$,
           Harald Kuntschner$^4$,
\newauthor Bryan W.\ Miller$^5$, G.\ Monnet$^6$, Reynier F.\ Peletier$^7$,
           E.K.\ Verolme$^1$ \\
$1$Sterrewacht Leiden, Niels Bohrweg 2, 2333 CA, Leiden, The Netherlands \\
$2$CRAL-Observatoire, 9 Avenue Charles--Andr\'e, 69230 Saint-Genis-Laval,
   France\\
$3$Department of Astronomy, Columbia University, 538 West 120th Street,
   New York, NY 10027, USA \\
$4$Physics Department, University of Durham, South Road, Durham
   DH13LE, United Kingdom \\
$5$Gemini Observatory, Casilla 603, La Serena, Chile\\
$6$European Southern Observatory, Karl-Schwarzschild Strasse 2,
   D-85748 Garching, Germany\\
$7$Department of Physics and Astronomy, University of
   Nottingham, University Park, Nottingham NG7 2RD, United Kingdom }

\date{Accepted $\ldots$
      Received $\ldots$;
      in original form $\ldots$ $\qquad$ }

\begin{document}
\maketitle

\begin{abstract}
Early results are reported from the \sauron\ survey of the kinematics
and stellar populations of a representative sample of nearby E, S0 and
Sa galaxies. The survey is aimed at determining the intrinsic shape of
the galaxies, their orbital structure, the mass-to-light ratio as a
function of radius, the age and metallicity of the stellar
populations, and the frequency of kinematically decoupled cores and
nuclear black holes.  The construction of the representative sample is
described, and its properties are illustrated. A comparison with
long-slit spectroscopic data establishes that the \sauron\
measurements are comparable to, or better than, the highest-quality
determinations.  Comparisons are presented for NGC 3384 and NGC 4365
where stellar velocities and velocity dispersions are determined to a
precision of 6~\kms, and the $h_3$ and $h_4$ parameters of the
line-of-sight velocity distribution to a precision of better than
0.02. Extraction of accurate gas emission-line intensities, velocities
and line widths from the datacubes is illustrated for NGC 5813.
Comparisons with published line-strengths for NGC 3384 and NGC 5813
reveal uncertainties of $\lta 0.1$\AA\ on the measurements of the
H$\beta$, Mg~{\it b} and Fe5270 indices.  Integral-field mapping
uniquely connects measurements of the kinematics and stellar
populations to the galaxy morphology. The maps presented here
illustrate the rich stellar kinematics, gaseous kinematics, and
line-strength distributions of early-type galaxies. The results
include the discovery of a thin, edge-on, disk in NGC~3623, confirm
the axisymmetric shape of the central region of M32, illustrate the
LINER nucleus and surrounding counter-rotating star-forming ring in
NGC~7742, and suggest a uniform stellar population in the decoupled
core galaxy NGC~5813.\looseness=-2
\end{abstract}

\begin{keywords}
galaxies: elliptical and lenticular ---
galaxies: individual
          (M32, NGC 3384, NGC 3623, NGC 4365, NGC 5813, NGC 7742) ---
galaxies: kinematics and dynamics ---
galaxies: spirals ---
galaxies: stellar content ---
integral-field spectroscopy
\end{keywords}

\section{Introduction}
\label{sec:intro}
The physical properties of early-type galaxies correlate with
luminosity and environment. The morphology-density relation (Dressler
1980) shows that elliptical and lenticular galaxies are much more
common in clusters than they are in regions of lower local density.
Giant ellipticals are red, have a high metal content, often have boxy
isophotes, are supported by anisotropic velocity distributions, and
have triaxial figures (e.g., de Zeeuw \& Franx 1991) whereas
lower-luminosity systems are bluer, less metal-rich, have disky
isophotes, are flattened by rotation, and may well have nearly-oblate
shapes (Davies et al.\ 1983; Bender \& Nieto 1990; de Zeeuw \& Carollo
1996; Faber et al.\ 1997).

Observations with the Hubble Space Telescope have re-enforced the
connection between the global and core properties of early-type
galaxies. The nuclei of elliptical and lenticular galaxies, as well as
the bulges of early-type spirals, often contain stellar and/or gaseous
disks, possibly associated with a massive central black hole, as well
as unresolved nuclear spikes, kinematically decoupled structures, and
puzzling asymmetries. These {\it nuclear} properties loosely correlate
with the {\it global} structure of the galaxies. For example, while
the luminosity profiles of elliptical galaxies approach a power-law
$\rho (r)\propto r^{-\gamma}$ at small radii $r$, giant ellipticals
($M_B\!  <\! -20.5$) have {\it shallow} cusps (mean $\gamma
\approx0.8$) and low-luminosity ellipticals ($M_B\! > \!-20.5$) have
{\it steep} cusps (mean $\gamma \approx1.9$) (Jaffe et al.\ 1994;
Faber et al.\ 1997). At intermediate luminosities, both types of
profile occur (Carollo et al.\ 1997; Rest et al.\ 2001). Furthermore,
it appears that the mass of the central black hole correlates with the
velocity dispersion of the host galaxy (Ferrarese \& Merritt 2000;
Gebhardt et al.\ 2000b).\looseness=-2

At present, it is unclear to what extent the observed properties of
these early-type galaxies, and the correlations between them, were
acquired at the epoch of galaxy formation or result from subsequent
dynamical evolution, possibly driven by a central black hole (Franx,
Illingworth \& de Zeeuw 1991; de Zeeuw 1996; Faber et al.\ 1997;
Merritt 1999; Bak \& Statler 2000). Key questions in this area
include: What is the distribution of intrinsic shapes, tumbling
speeds, and internal orbital structure of early-type galaxies? How do
these depend on total luminosity and environment? What is the shape
and extent of the dark halo? What is the dynamical importance of the
central black hole? What is the distribution of metals, and what is
the relation between the kinematics of stars (and gas), the local
metal enrichment, and the star formation history?

Progress towards answering these questions requires a systematic
investigation of the kinematics and line-strengths of a representative
sample of early-type systems. The intrinsic shape, internal orbital
structure, and the radial dependence of the mass-to-light ratio $M/L$
are constrained by the line-of-sight velocity distribution of the
stars (e.g., van der Marel \& Franx 1993; Gerhard 1993) and by
the gas kinematics (e.g., Cretton, Rix \& de Zeeuw 2000). The
absorption line-strengths can be used to infer the luminosity-weighted
age and metallicity of the stellar populations (Gonz\'alez 1993;
Carollo, Danziger \& Buson 1993; Davies, Sadler \& Peletier 1993;
Worthey 1994). Detailed studies of individual galaxies are valuable,
but need to be complemented with a systematic investigation of the
internal structure as a function of global properties, such as total
luminosity and Hubble type, and environment.\looseness=-2

Studies of the kinematics and line-strength distributions of galaxies
have traditionally relied on long-slit spectroscopy along at most a
few position angles (e.g., Davies \& Birkinshaw 1988; Franx,
Illingworth \& Heckman 1989; Bender, Saglia \& Gerhard 1994; Surma \&
Bender 1995; Statler \& Smecker--Hane 1999).  This is insufficient to
map the rich internal structure of early-type galaxies (Statler 1991,
1994; Arnold, de Zeeuw \& Hunter 1994). Constructing full
two-dimensional maps for a significant sample of galaxies with
long-slit spectroscopy requires a prohibitive amount of telescope
time. For this reason, we built \sauron, a panoramic integral-field
spectrograph optimized for studies of the large-scale kinematics and
stellar populations of galaxies (Bacon et al.\ 2001, hereafter Paper
I). \sauron\ is based on the same principles as the prototype
integral-field spectrograph \tiger, built for the CFHT, and its
successor \oasis\ (Bacon et al.\ 1995, 2000).  In its low-resolution
(LR) mode, \sauron\ has a $33''\times41''$ field-of-view sampled with
$0\farcs94\times0\farcs94$ pixels, 100\% coverage, high throughput,
and a spectral resolution of 3.6\AA FWHM. In high-resolution (HR)
mode, the spectral resolution is 2.8\AA\ and the field of view is
$9''\times11''$, sampled at $0\farcs27\times0\farcs27$.  \sauron\
provides 1577 spectra in one exposure (wavelength range 4810--5350\AA,
see also \S\ref{sec:n3384-lines}), 146 of which are used for
simultaneous sky subtraction $1\farcm9$ away from the main field.
Stellar kinematical information is derived from the Mg{\it b} triplet
and the Fe lines. The [OIII] and H$\beta$ emission lines allow
derivation of the morphology and kinematics of the ionized gas, and
provide information on its ionization state. The Mg{\it b}, H$\beta$,
and Fe5270 absorption lines are sensitive to the age and metallicity
of the stellar populations.  Paper I describes the design and
construction of \sauron, and the extensive data reduction software
developed. \sauron\ was commissioned on the William Herschel Telescope
(WHT) in February 1999 (de Zeeuw et al.\ 2000).

We are using \sauron\ on the WHT to observe a representative sample of
nearby ellipticals, lenticulars, and early-type spiral bulges, as well
as some objects with known peculiar kinematics (e.g., Davies et al.\
2001). We will combine the \sauron\ observations with high-spatial
resolution spectroscopy of the nuclei, and interpret them through
dynamical and stellar population modeling. Here we present selected
observations which serve to establish the spectrograph's scientific
integrity, and provide early results from our systematic survey.
\S\ref{sec:project} describes the definition of the sample.
\S\ref{sec:stellar-kin} presents stellar kinematic measurements
showing evidence for embedded disks and decoupled cores.
\S\ref{sec:gas-kin} is devoted to measurements of the kinematics and
line-ratios of extended emission-line gas, and
\S\ref{sec:line-strengths} presents absorption line-strength
measurements. In each of these three sections, we compare the \sauron\
results with previous long-slit measurements, and show that \sauron\
is efficient and accurate.  \S\ref{sec:conclusions} summarises the
results and outlines the next steps.

\begin{figure*}
\centerline{\psfig{file=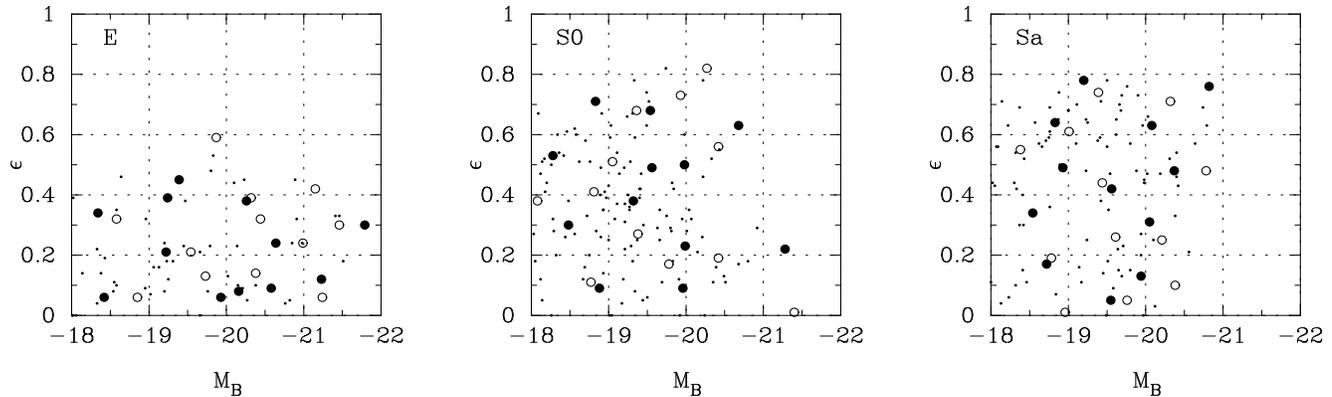,width=17.5cm}}
\caption{Distribution of 72 E, S0, and Sa galaxies in the \sauron\
representative sample, in the plane of ellipticity $\epsilon$ versus
absolute blue magnitude $M_B$. Filled circles: galaxies in the Virgo
cluster, or the Leo and Coma I groups (`cluster' objects). Open
circles: other galaxies (`field' objects). Small dots: non-selected
255 galaxies of the complete sample. }
\label{fig:sample}
\end{figure*}

\section{The SAURON sample}
\label{sec:project}

Observing any complete sample which spans a wide range of global
properties is costly in telescope time, even with \sauron. We have
therefore constructed a {\it representative} sample of nearby
early-type galaxies, as free of biases as possible while ensuring the
existence of complementary data.  \looseness=-2

\subsection{Selection}
\label{sec:sample}

We first compiled a complete list of elliptical galaxies, lenticular
galaxies, and spiral bulges for which \sauron\ can measure the stellar
kinematics.  Given the specifications of the instrument when mounted
on the WHT (Paper~I), this leads to the following constraints:
$-6^\circ \leq \delta \leq 64^\circ$ (to limit the zenith distance and
therefore the instrumental flexure), $cz \leq 3000$~\kms\ (to ensure
that all the lines of interest are in the spectral band), and $M_B
\leq -18$ (so that $\sigma \gta 75$~\kms\ and velocity dispersions can
be measured).  We further restricted the objects to $|b| \geq
15^\circ$ to avoid crowded fields and large Galactic extinctions. All
parameters except $M_B$ were taken from the Lyon/Meudon Extragalactic
Database (LEDA; see Paturel et al.\ 1997) and checked for consistency
with the RC3 (de Vaucouleurs et al.\ 1991) through VizieR (Ochsenbein,
Bauer \& Marcout 2000).\looseness=-2

We derived absolute magnitudes following the prescription in LEDA,
using the listed heliocentric velocities and apparent magnitudes. We
adopted a Virgocentric flow model with $v_{\rm virgo} = 170$ \kms,
$H_0 = 75$ \kms, and the correction to the Local Group centroid of
Yahil, Tammann \& Sandage (1977). For galaxies in the Virgo cluster,
the Coma~I cloud, and the Leo~I group, which we refer to as `cluster'
galaxies, we adopted common distances based on the mean heliocentric
velocity of each group, taken from Mould et al.\ (1993).  The
distances derived, respectively 16.3, 13.7, and 10.7~Mpc (31.04, 30.68
and 30.14 mag), are in good agreement with those derived from other
measurements (see, e.g., Ferrarese et al.\ 2000). For galaxies outside
these three associations, which we refer to as `field' galaxies, we
used individual distances. We took cluster and group membership
primarily from Tully (1988) and Garcia (1993), but also from other
sources (e.g., Turner \& Gott 1976; Huchra \& Geller 1982; Geller \&
Huchra 1983; Faber et al.\ 1989).

The list of galaxies obtained with the above criteria contains 327
objects. We divided these into six categories by first separating
`cluster' and `field' galaxies, and then splitting each of these groups in
E, S0, and Sa galaxies, based on LEDA's morphological type (E: $T \leq
-3.0$, S0: $-3.0 < T \leq -0.5$, Sa: $-0.5 < T \leq 1.5$).  This
results in 29 E, 51 S0 and 37 Sa galaxies in the `cluster' environment,
and 47 E, 86 S0 and 77 Sa galaxies in the `field'. We selected a {\em
representative} sample of objects from this complete list by
populating the six ellipticity versus absolute magnitude planes nearly
uniformly, using bins of 1 magnitude in luminosity and 0.2 in
ellipticity (where $\epsilon$ is derived from the axial ratio at the
25~mag~arcsec$^{-2}$ level in $B$, as reported in LEDA; for the
spirals it effectively gives the inclination of the disk). This choice
of bin size is appropriate given the range of absolute magnitude we
want to cover, and the radial ellipticity variations observed in
early-type galaxies. The result is 36 `cluster' galaxies (12 E, 12 S0,
and 12 Sa), and 36 `field' galaxies (12 E, 12 S0, and 12 Sa), as
illustrated in Figure~\ref{fig:sample}. A sample of this size is large
enough to be sub-divided by morphology, environment and luminosity,
yet small enough that complete mapping with \sauron\ is possible. The
72 galaxies correspond to 22\% of the complete sample and, as can be
seen from Figure~\ref{fig:sample}, remain representative of it.
Tables~\ref{tab:props-cluster} and \ref{tab:props-field} list the 72
galaxies and summarise some of their properties.\looseness=-2

\begin{figure}
\centerline{\psfig{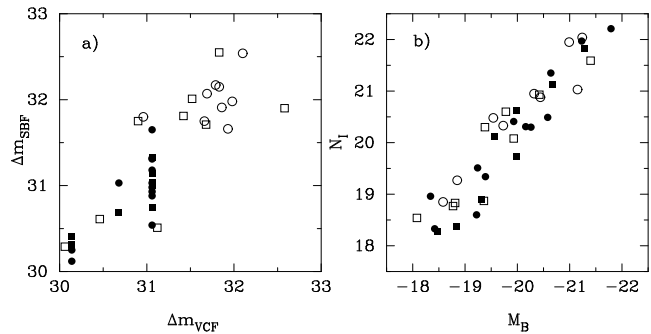}}
\caption{Comparison of distances and absolute luminosities for the
galaxies in the \sauron\ sample for which
surface-brightness-fluctuation measurements were reported by Tonry et
al.\ (2001).  a) SBF distance moduli $\Delta m_{\rm SBF}$ versus
Virgocentric distance moduli $\Delta m_{\rm VCF}$. b) SBF fluctuation
star counts $\overline{N}_I$ versus absolute magnitudes $M_B$.
Circles: elliptical galaxies. Squares: lenticular galaxies.  Filled
symbols: `cluster' objects. Open symbols: `field' objects. No spiral
galaxies are in common between the two sets.\looseness=-2}
\label{fig:sample-sbf}
\end{figure}

Figure~\ref{fig:sample-sbf} compares distances and absolute
luminosities for the 38 sample galaxies in common with the $I$-band
surface brightness fluctuation (SBF) survey of Tonry et al.\ (2001).
$\overline{N}_I$ is a distance-independent measure of absolute
luminosity. Our distances are in very good agreement with the SBF
distances ($\langle |\Delta m_{\rm SBF}-\Delta m_{\rm
VCF}|\rangle=0.13$), and our $M_B$ correlates strongly with
$\overline{N}_I$. Only seven galaxies have distances differing by more
than 0.5~mag (half a bin) from the mean offset. Despite a simplistic
approach, our distance estimates are thus accurate enough for the
chosen bin size.

\begin{figure*}
\centerline{\psfig{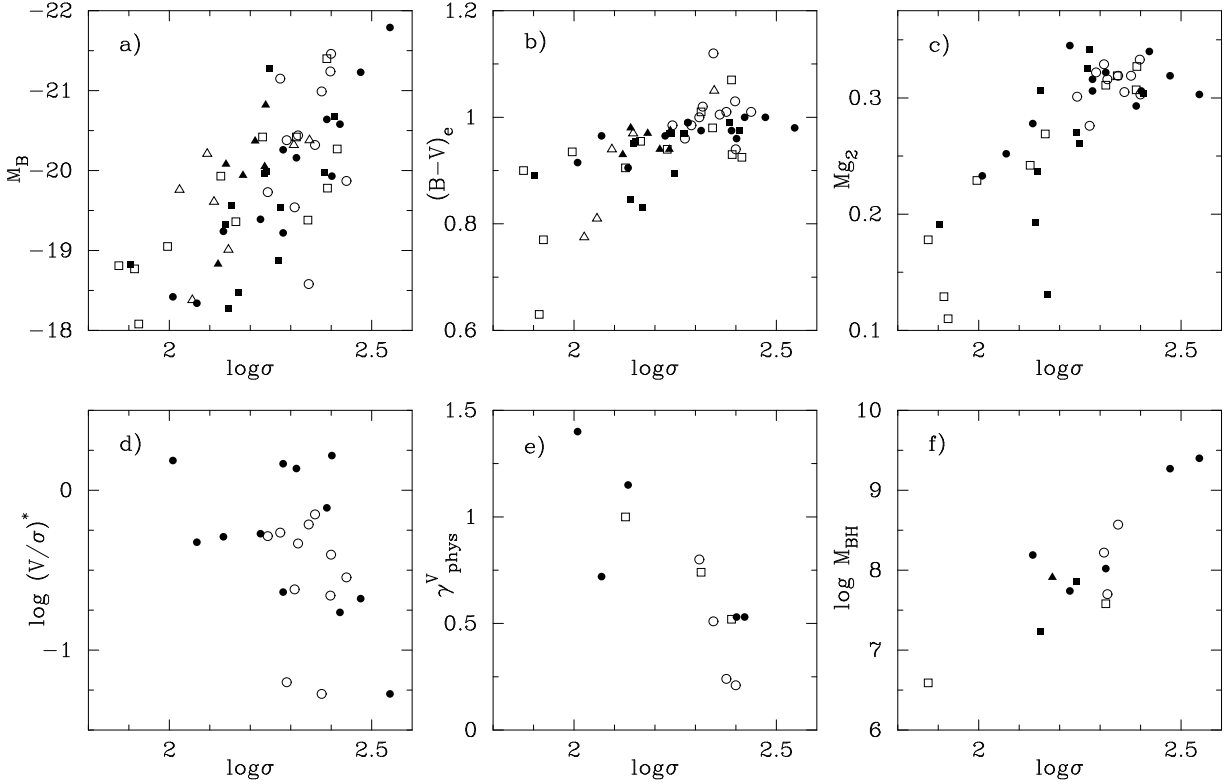}}
\caption{Properties of the galaxies in the \sauron\ representative
sample. Circles: elliptical galaxies. Squares: lenticular
galaxies. Triangles: early-type spiral bulges. Filled symbols: `cluster'
objects. Open symbols: `field' objects.  Plotted versus central velocity
dispersion $\sigma$ are a) absolute blue magnitude $M_B$, b) effective
$B -V$ colour, c) central Mg$_2$ line-strength, d) measure of
rotational support $(V /\sigma)^*$, e) central cusp-slope
$\gamma^V_{\rm phys}$ of the surface-brightness profile and f) black hole
mass $M_{\rm BH}$.  The data are taken from the literature (see text,
and Tables~\ref{tab:props-cluster} and \ref{tab:props-field}) and
error bars are not shown. Most quantities are not available for all
galaxies. }
\label{fig:sample-properties}
\end{figure*}

The complete list of galaxies is unbiased, but we did consider
additional criteria when selecting the representative sample. In cases
where more then one galaxy could be used to fill a given bin, we
favored the one for which high spatial resolution imaging and
kinematics are available, primarily Hubble Space Telescope (HST) {\tt
WFPC2} and {\tt STIS} observations from the HST archive. Such
observations are essential to constrain the mass distribution in the
nuclei of the galaxies.  Similarly, we preferentially chose galaxies
for which kinematics at large radii is available, for example through
long-slit spectroscopy or HI observations. These data can be used to
evaluate the importance of dark matter in the outer parts of the
galaxies.  In the case of S0 and Sa galaxies, we also tried to obtain
a representative mix of barred and unbarred systems. These choices
introduce a bias in the selection, but this bias is hard to
quantify. The ground-based observations were primarily compiled using
the Hypercat catalogue (e.g., Prugniel et al.\ 1997) and derive from a
range of programs with different goals, but a large fraction of them
were carried out as part of statistically complete surveys. Thus,
overall, we believe that the bias introduced is minimal, and that the
advantages of working with a carefully crafted sample far outweigh the
disadvantages.

\subsection{Properties of the sample galaxies}
\label{sec:properties}

Figure~\ref{fig:sample-properties} illustrates that the galaxies in
our representative sample cover a large range of global and nuclear
properties (listed in Tables~\ref{tab:props-cluster} and
\ref{tab:props-field}). Panel {\it a} presents the Faber--Jackson
(1976) relation between total luminosity (absolute blue magnitude
$M_B$; see \S\ref{sec:sample}) and central velocity dispersion
$\sigma$ (LEDA). The galaxies cover a range of $\approx$50 in
luminosity and a factor of five in velocity dispersion.  Panel {\it b}
plots the effective $B-V$ colour (LEDA) versus $\sigma$, and panel
{\it c} plots the central value of Mg$_2$ (Hypercat) versus
$\sigma$. These panels demonstrate that large galaxies are
systematically redder and have stronger lines.  Panel {\it d} shows
the ellipticity-normalised anisotropy parameter $(V/\sigma)^*$ (i.e.\
$V_{\rm max}/\sigma$ divided by the value $\sqrt{\epsilon/1-\epsilon}$
for the oblate isotropic rotator of the same ellipticity as the
galaxy, where $V_{\rm max}$ is the maximum of the observed stellar
mean streaming velocity, see Davies et al.\ 1983), plotted against the
central velocity dispersion for the elliptical galaxies (LEDA).  The
ellipticities used are those from column 10 of
Tables~\ref{tab:props-cluster} and \ref{tab:props-field}, and are not
necessarily applicable to the domain of the galaxy where the stellar
rotation velocity was measured. The values for $V_{\rm max}$ are the
averages of the maximum stellar $V$ measured along the major axis for
sources listed in LEDA (column 12 of Tables~\ref{tab:props-cluster}
and \ref{tab:props-field}). Upper limits are included as
measurements. Discrepancies in the catalog entries were resolved by
examining the original data. The elliptical galaxies in the sample
span a wide range of anisotropy.  About two dozen galaxies have
well-determined nuclear surface brightness profiles derived from HST
imaging (Lauer et al.\ 1995; Byun et al.\ 1996; Carollo et al.\ 1997;
Rest et al.\ 2001). Panel {\it e} shows the average logarithmic slope
$\gamma^V_{\rm phys}$ of the deprojected luminosity profile within
10--50~pc, as derived by Carollo et al.\ (1997, Tables 5 and 8) for 12
of our objects.  They display a significant scatter, but
lower-dispersion galaxies generally have steeper central cusps. A
similar number of objects in our sample also have published values for
the mass of a central black hole, based on high-resolution absorption
or emission-line spectroscopy and (in most cases) axisymmetric
three-integral dynamical modeling (Bower et al.\ 1998, 2001; Gebhardt
et al.\ 2000a, b; Macchetto et al.\ 1997; Sarzi et al.\ 2001). As
Panel {\it f} shows, they cover three orders of magnitude in black
hole mass.

\begin{figure*}
\centerline{\psfig{file=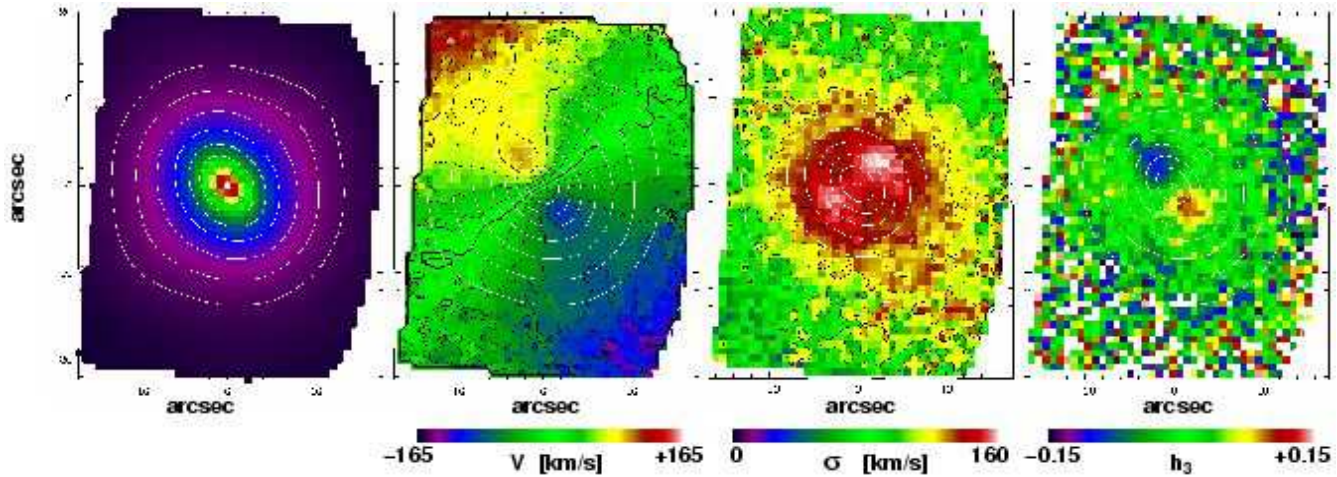,width=17.7cm}}
\caption{\sauron\ measurements of the SB0 galaxy NGC 3384, based on a
single pointing consisting of four 1800~s exposures. The field-of-view
is $33''\times41''$, and the effective spatial sampling is
$0\farcs8\times0\farcs8$. From left to right: a) reconstructed total
intensity $I$. b) stellar mean velocity $V$, with the zero-velocity
contour indicated by the thick curve. c) stellar velocity dispersion
$\sigma$. d) Gauss-Hermite moment $h_3$ which measures the skewness of
the line-of-sight velocity distribution. The Gauss--Hermite moment
$h_4$ was measured as well, but varies little over the field, and is
not shown. Contours of the reconstructed total intensity are
superimposed (in 0.25 mag~arcsec$^{-2}$ steps). }
\label{fig:n3384-obs-kin}
\end{figure*}

\section{Stellar kinematics}
\label{sec:stellar-kin}

Our strategy is to map galaxies out to one effective radius $R_e$,
which for nearly half the sample requires only one \sauron\
pointing. For large galaxies, mosaics of two or three pointings reach
0.5$R_e$, except for M84 and M87 where two additional pairs of
flanking fields are needed to reach this minimum radius. Each two hour
pointing is typically split into four exposures of 1800 s each,
dithered by one lenslet.  We reduce and merge the raw \sauron\
exposures as described in Paper I, and use the individual
wavelength-calibrated and extracted spectra to derive maps of the
stellar kinematics using the FCQ method (Bender 1990) and an optimal
stellar template (Paper I, \S 5.2). This provides the mean stellar
velocity $V$, the velocity dispersion $\sigma$, as well as the
Gauss-Hermite moments $h_3$ and $h_4$, which quantify the first-order
asymmetric and symmetric deviations of the line-of-sight velocity
distribution from a Gaussian form (van der Marel \& Franx 1993;
Gerhard 1993).

We estimated the errors on the kinematic parameters by means of
extensive Monte-Carlo tests: for each galaxy spectrum and its
associated set of kinematic parameters, an initial noise-free galaxy
spectrum was estimated by convolving the template spectrum with the
corresponding analytic line-of-sight velocity distribution.
Extraction of a new set of parameters after inclusion of a realization
of the noise at the given signal-to-noise constitutes a random
realization out of the statistical distribution of these parameters,
and the errors on the kinematic parameters are estimated from the
variance of 100 realizations of this distribution.

\sauron\ observations of the kinematics of stars and gas in the E6
galaxy NGC~3377 were presented in Figure 12 of Paper I. Kinematic and
line-strength maps of NGC~4365 were published by Davies et al.\
(2001). In this section we present the \sauron\ kinematics of NGC~3384
observed with the LR mode, and compare the results with high-quality
long-slit kinematics from the literature. We then discuss NGC 3623 and
NGC 4365, both of which display significant structure in the observed
kinematics which is not captured by long-slit spectroscopy along a few
position angles. We conclude by presenting kinematic maps of M32 taken
with the HR mode.\looseness=-2

\subsection{LR mode observations of NGC 3384}
\label{sec:n3384-kin}

NGC~3384 is a large early-type spiral in the Leo~I group, classified
as SB0$^-$(s) in the RC3 ($M_B=-19.56$, $T=-2.6$ in LEDA). It forms a
triple on the sky with NGC~3379 and NGC~3389, but there is only
marginal evidence for interactions in the form of a faint spiral or
tidal arm (Malin 1984) and the existence of a large HI ring around the
triple (Schneider 1985). Barbon, Capaccioli \& Tarenghi (1975)
described the three main components of NGC~3384: a bright $R^{1/4}$
central region ($r\lta20\arcsec$), a lens ($r\lta160\arcsec$), and an
outer exponential disk ($r\gta160\arcsec$). The light distribution in
the central $\approx$$20\arcsec$ is complex.  The inner isophotes are
elongated along the major axis, suggesting an embedded disk, but
beyond 10$''$ the elongation is along the minor-axis, possibly due to
a faint polar ring or bar (Davoust et al.\ 1984; Whitmore et al.\
1990; Busarello et al.\ 1996). At much larger radii the isophotes are
off-centred.  NGC~3384 shows no emission lines (Ho, Filippenko \&
Sargent 1997), remains undetected in HI, CO, radio continuum, and
X-ray (e.g.\ Roberts et al.\ 1991), but has IRAS 12 and 100~$\mu$m
fluxes (Knapp et al.\ 1989).\looseness=-2

\begin{figure*}
\centerline{\psfig{file=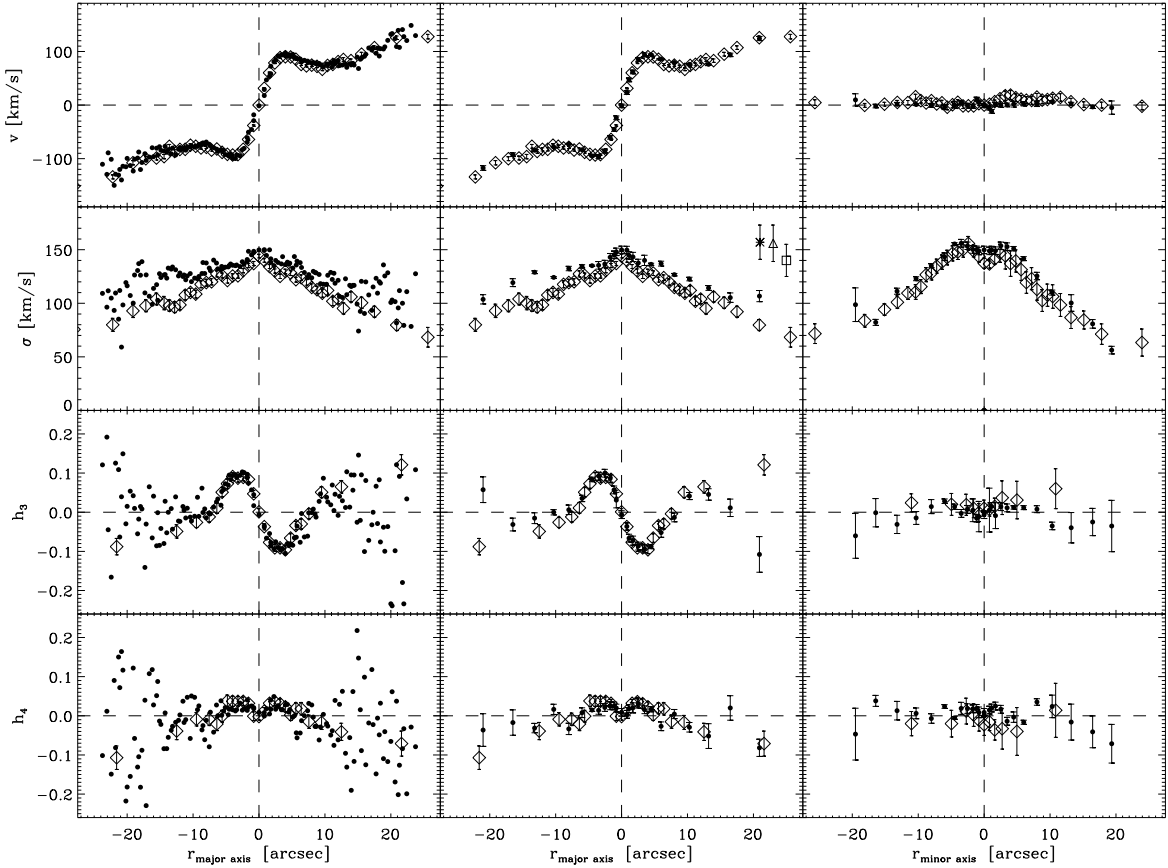,width=15.5truecm}}
\caption{Comparison of the {\tt SAURON} kinematics of NGC 3384 (dots)
with the long-slit measurements of Fisher (1997, diamonds, $2''$
slit).  Top to bottom: mean velocity $V$, velocity dispersion
$\sigma$, and Gauss-Hermite moments $h_3$ and $h_4$. The left panels
show the comparison for the major axis (PA $=53^\circ$) with the
\sauron\ measurements for all lenslets whose centres fall inside a
$2''$ wide `slit' along the major axis. Individual error bars are not
shown. The middle panels show the same comparison, with the \sauron\
data logarithmically binned in radius, and the errors on the mean
included. Central velocity dispersion measurements by Tonry \& Davis
(1981, asterisk, $3''\times 12''$ aperture), Dalle Ore et al.\ (1991,
triangle, $1\farcs5\times4\farcs0$ aperture) and Neistein et al.\
(1999, square, $2''$ slit) are shown in the top right corner of the
second panel from the top (offset in radius).  The right panels show
the similar comparison for the minor axis
(PA~$=143^\circ$).\looseness=-2 }
\label{fig:n3384-kin-comp}
\end{figure*}

We observed NGC~3384 with \sauron\ in LR mode on the night of April 4,
2000. We obtained $4\times1800$~s on one field centred on the galaxy,
which covers most of the bulge. The seeing was $2\farcs5$.
Figure~\ref{fig:n3384-obs-kin} displays the resulting integrated
intensity, the mean stellar velocity $V$ and velocity dispersion
$\sigma$, and the Gauss-Hermite moments $h_3$ and $h_4$.\looseness=-2

Figure~\ref{fig:n3384-obs-kin}{\it a} illustrates a key advantage of
integral-field spectroscopy over traditional aperture and long-slit
spectra: by integrating the flux in each spectrum, the two-dimensional
surface brightness distribution of the galaxy is recovered. As this is
derived from the same data that are used to obtain the kinematics and
line-strengths, there is no doubt about the relative location of these
measurements. A comparison with an HST/F555W image taken from the HST
archive shows excellent agreement, after taking into account the
$2\farcs5$ seeing and the spatial sampling of \sauron.\looseness=-2

\begin{figure*}
\centerline{\psfig{file=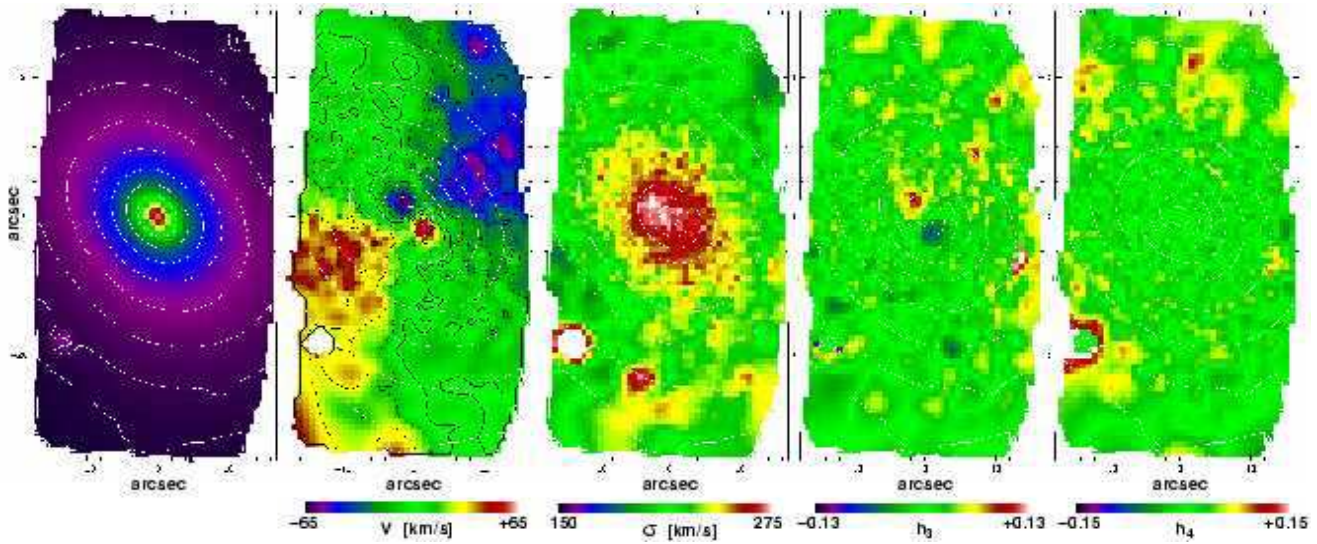,width=17.5cm,clip=}}
\caption{\sauron\ observations of the E3 galaxy NGC~4365. From left to
right: a) reconstructed total intensity $I$. b) stellar mean velocity
$V$, with the zero-velocity contour indicated by the thick solid
curve. c) velocity dispersion $\sigma$. d) $h_3$. e) $h_4$. Contours
of $I$ are superimposed (0.25 mag~arcsec$^{-2}$ per step). The maps
are based on two pointings of $4\times 1800$~s each, sampled with
$0\farcs8\times0\farcs8$ pixels and a $\approx$$20''$ overlap. The
total field of view is $33''\times63''$. Maps of Mg{\it b} and
H$\beta$ based on the same observations are shown in Davies et al.\
(2001).\looseness=-2 }
\label{fig:n4365-obs}
\end{figure*}

Figure~\ref{fig:n3384-obs-kin}{\it b} shows that the bulge of NGC~3384
exhibits a regular rotation field. The mean velocities increase
steeply along the major axis, reaching $\approx$110~\kms\ at $r\approx
4''$, then decrease slightly, after which they rise again.  No
velocity gradient is observed along the minor axis, indicating that
the kinematics in the inner region of NGC 3384 is consistent with
axisymmetry.  The velocity dispersion map
(Fig.~\ref{fig:n3384-obs-kin}{\it c}) shows a symmetric dumb-bell
structure with a central depression.  Fisher (1997) had already
observed the central dip in the minor axis velocity dispersion
profile, together with an abrupt change in the $h_3$ gradient at
$r\approx4\arcsec$ along the major axis (cf.\
Fig~\ref{fig:n3384-obs-kin}{\it d}).  These results strongly suggest
the presence of a rapidly rotating inclined inner stellar
disk.\looseness=-2

Busarello et al.\ (1996) reported long-slit measurements of $V$ and
$\sigma$ along three position angles.  Fisher's (1997) higher
signal-to-noise measurements of $V$, $\sigma$, $h_3$ and $h_4$ along
the major and the minor axis (position angles 53$^\circ$ and
143$^\circ$) have a spectral resolution very similar to that of
\sauron. Figure~\ref{fig:n3384-kin-comp} compares his results with
ours. The column of four panels on the left shows the direct
comparison with the \sauron\ measurements in all the individual
lenslets whose centres fall inside the slit used by Fisher.  The
middle and right columns of panels show the comparison with the
\sauron\ data now logarithmically binned in radius to increase the
signal-to-noise. The agreement is good and shows that a two hour
exposure provides reliable kinematics over the entire \sauron\ field.
Re-sampling our data to the spatial sampling by Fisher results in RMS
deviations of 7 km/s in $V$, 8 km/s in $\sigma$, 0.02 in $h_3$ and
0.02 in $h_4$. We do not find systematic offsets for $V$ and
$h_4$. For $\sigma$ we find mean offsets of 14 km/s and 7 km/s ({\tt
SAURON} -- Fisher) for the major and minor axis, respectively. Our
slightly larger $\sigma$'s are, however, consistent with the central
measurements of Tonry \& Davis (1981, asterisk), Dalle Ore et al.\
(1991, open triangle) and Neistein et al.\ (1999, open square), as
shown in Figure~\ref{fig:n3384-kin-comp}. The $h_3$ values agree well,
except beyond 15$''$ where the binned \sauron\ measurements indicate a
return to $h_3$ values with the opposite sign to that of $V$, as seen
in many elliptical galaxies (e.g., Bender, Saglia \& Gerhard
1994). Taking into account the RMS variation in Fisher's data, we
conclude that the RMS errors in the \sauron\ measurements are $\Delta
V \approx 6$ \kms, $\Delta \sigma \approx 6$ \kms, $\Delta h_3 \approx
0.015$ and $\Delta h_4 \approx 0.015$.

Gebhardt et al.\ (2000b) reported the presence of a central black hole
of $(1.7\pm0.8)\times 10^7 M_\odot$ in the nucleus of NGC~3384, based
on \stis\ absorption-line spectroscopy. The spatial resolution of the
\sauron\ maps (and of Fisher's long-slit data) is insufficient to
resolve the increase in dispersion inside the radius of influence of
such a black hole. A dynamical model for NGC~3384 which combines the
\stis\ results with \oasis\ and \sauron\ data will provide an accurate
determination of the intrinsic shape and internal velocity
distribution of the bulge and disk of NGC 3384, and should also
increase the accuracy of the black hole mass determination.

\subsection{The decoupled core of NGC 4365}
\label{sec:n4365}

NGC~4365 is an E3 galaxy located in the Virgo W cloud just beyond the
centre of the Virgo cluster. Surma \& Bender (1995) showed that the
main body of NGC~4365 rotates around its projected major axis and
that its core rotates around the minor axis (cf.\ Bender 1988; Wagner,
Bender \& M\"ollenhoff 1988). The central luminosity profile has a
shallow cusp, and the nucleus contains a blue central point source
(Carollo et al.\ 1997).

We observed NGC~4365 with \sauron\ on the nights of 29 \& 30 March
2000. Two pointings of two hours each, overlapping by
$\approx$20\arcsec\/ on the nucleus, covered a total region of
33\arcsec$\times$63\arcsec.  The average seeing during the
observations was $\approx$$1\farcs6$ (see Davies et al.\ 2001 for
further details).  Figure~\ref{fig:n4365-obs} presents maps of the
reconstructed total intensity, the mean stellar velocity $V$ and
velocity dispersion $\sigma$, and the Gauss--Hermite moments $h_3$ and
$h_4$.

\begin{figure}
\centerline{\psfig{file=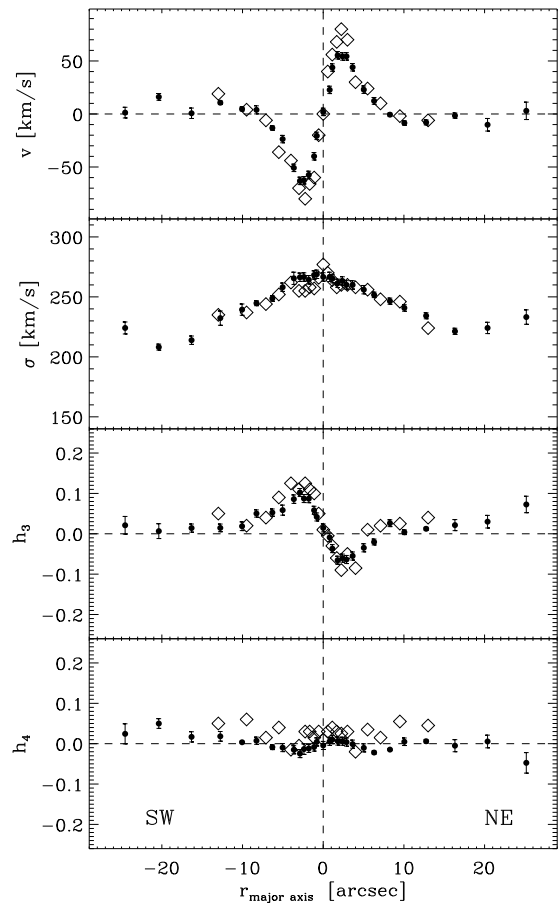,width=7.7cm,clip=}}
\caption{Comparison of the logarithmically binned \sauron\ kinematics
along the major axis of NGC~4365 (dots) with the (binned) long-slit
measurements of Surma \& Bender (1995, diamonds). From top to bottom:
$V$, $\sigma$, $h_3$ and $h_4$. }
\label{fig:n4365-comparison}
\end{figure}

\begin{figure*}
\centerline{\psfig{file=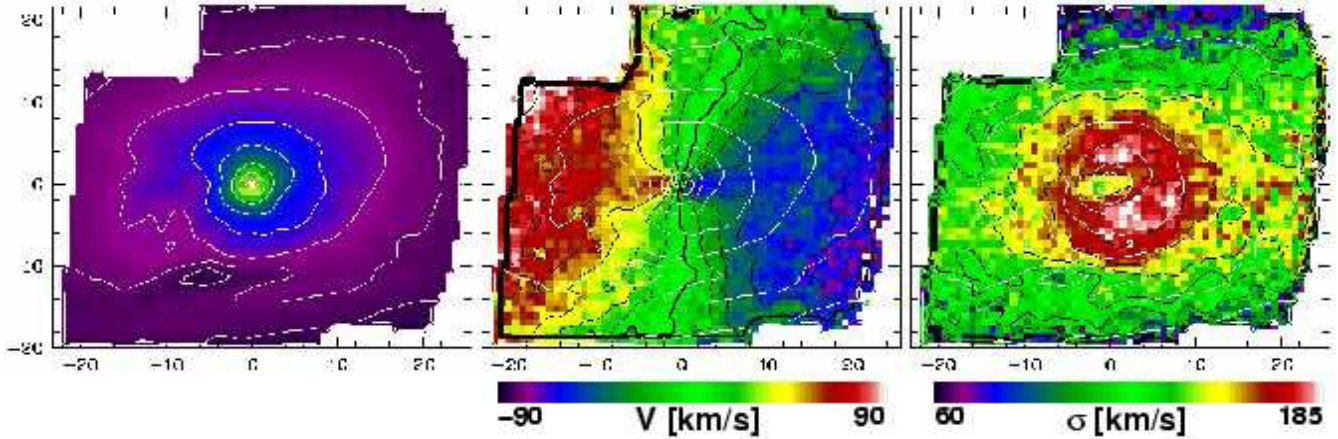,width=17.7truecm,clip=}}
\caption{\sauron\ measurements of NGC 3623 in the Leo triplet, based
on two pointings overlapping by $\approx$$15''$, and exposed for
$4\times1800$~s each.  The field-of-view is $40''\times 50''$ and the
spatial sampling is $0\farcs8\times0\farcs8$. From left to right: a)
reconstructed total intensity $I$, b) stellar mean velocity $V$, and
c) velocity dispersion $\sigma$. Contours of the reconstructed total
intensity are superimposed (steps of 0.25 mag arcsec$^{-2}$).  They
are significantly affected by a dust lane below the nucleus. }
\label{fig:n3623-obs}
\end{figure*}

The mean velocity field displayed in Figure~\ref{fig:n4365-obs}
clearly shows the rich kinematical structure of NGC~4365. As described
in Davies et al.\ (2001), the core measures $7''\times3''$ and rotates
around the minor axis, while the bulk of the galaxy rotates
approximately around the major axis.  The map of the $h_3$-moment
displays a similar two-component structure, with the values
anticorrelated with $V$. The Gauss-Hermite moment $h_4$ is nearly
zero, and shows very little variation over the field. The early
long-slit spectroscopy had revealed the global kinematic structure of
NGC~4365, but the \sauron\ maps make it possible to determine, for
example, the position angle of the rotation axis accurately. It is
aligned with the minor axis in the core, but misaligned from the major
axis by $12\pm2^\circ$ in the outer regions. This is a strong
indication of a triaxial intrinsic shape (cf.\ Statler 1991), possibly
containing a nearly axisymmetric core.  Construction of a triaxial
dynamical model for this object is in preparation.\looseness=-2

The individual \sauron\ measurements along the major axis can be
compared with the long-slit data from Surma \& Bender (1995), obtained
with a 1\farcs4 slit and binned radially to obtain sufficient
signal-to-noise. The original data could not be retrieved, so we
scanned the published figure. The agreement is excellent. The mean
velocity curves agree very well outside $3''$. Inside this radius
there are differences in the peak amplitude and the slope of the
central gradient, but they are consistent with the different values of
the seeing. Resampling our data to the spatial sampling used by Surma
\& Bender (1995) results in RMS deviations of 13 \kms\ in $V$, 6 \kms\
in $\sigma$, and 0.02 in $h_3$ and $h_4$. We ascribe the rather large
RMS deviation for $V$ to the different seeing. We do not find
significant systematic offsets for $V$, $\sigma$ and $h_3$.  For $h_4$
we find a systematic offset of 0.03 in the sense that our measurements
are smaller. This may be due to differences between the template stars
used in the two studies.\looseness=-2

\subsection{A central stellar disk in NGC 3623}
\label{sec:n3623}
NGC~3623 (M~65) is a bright, large, highly-inclined SAB(rs)a galaxy
located in the nearby Leo~I group ($M_B=-20.8$, $D_{25}=9\farcm8$,
$i\approx75^\circ$). It forms the Leo triplet together with NGC~3627 and
NGC~3628, and it is classified as a LINER (e.g.\ Ho, Filippenko \&
Sargent 1997). While there are strong indications that NGC~3627 and
NGC~3628 are interacting, NGC~3623 appears undisturbed (e.g., Chromey
et al.\ 1998 and references therein).  It is one of the galaxies
discussed by Hubble (1943) in his paper on the sense of rotation of
spiral arms and is included in the Hubble Atlas (Sandage 1961).

Surprisingly, despite its closeness, the kinematics of NGC~3623 has
received little attention. The only available absorption and
emission-line kinematics dates back to Burbidge, Burbidge \&
Prendergast (1961). All HI kinematic data are at low spatial
resolution (e.g., Krumm \& Salpeter 1979) or deal with the HI
distribution of the entire group (e.g., Rots 1978). Most spectroscopic
observations of NGC~3623 are concerned with emission-line ratios, to
probe the ionization mechanism of the gas and the nuclear activity
(from Burbidge \& Burbidge 1962 to Bresolin, Kennicutt \& Garnett
1999). The observations reported here offer the first view of the
stellar kinematics of the bulge of NGC~3623.

We observed NGC~3623 on 30 \& 31 March, 2000, and obtained a mosaic of
two fields with substantial overlap on the nucleus. Each field was
exposed for $4\times1800$~s and the resulting mosaic covers most of
the bulge. The seeing was $\approx$1$''$. The \sauron\ total
intensity, velocity, and velocity dispersion maps are shown in
Figure~\ref{fig:n3623-obs} (cf.\ Bureau et al.\ 2001). The main dust
lane is easily visible in the \sauron\ reconstructed image. It appears
as the dark band covering the bottom 25\% of the field (the position
angle of the major axis of the galaxy is roughly horizontal in
Fig.~\ref{fig:n3623-obs}). The velocity field shows a typical rotation
pattern, but also reveals twists in the kinematic minor axis and a
gradient along the minor axis (`minor-axis rotation'). This is not
unexpected since NGC~3623 is barred. There may be a contribution from
a warped stellar disk.  However, it is difficult to estimate how much
of this structure is affected by the dust extinction. One feature
which is certainly not due to dust is the central stellar disk,
clearly visible as a flattening of the isovelocity contours in the
centre of the galaxy. This disk is also easily seen in the velocity
dispersion map. While the dispersion increases steadily towards the
centre, roughly following the isophotes, there is a pronounced
elongated depression in the centre. This is the expected signature of
a flattened, cold component embedded in the bulge. The small thickness
observed suggests that the central disk is thin and located in the
equatorial plane of the galaxy (nearly edge-on).  Both the central
stellar disk and the large scale disk are likely coplanar. The $h_3$
map (not shown) shows a sign reversal at about 6--8\arcsec\ on the
major axis, confirming the size of the central disk apparent from
Fig~\ref{fig:n3623-obs}{\it b--c}.

\begin{figure*}
\centerline{\psfig{file=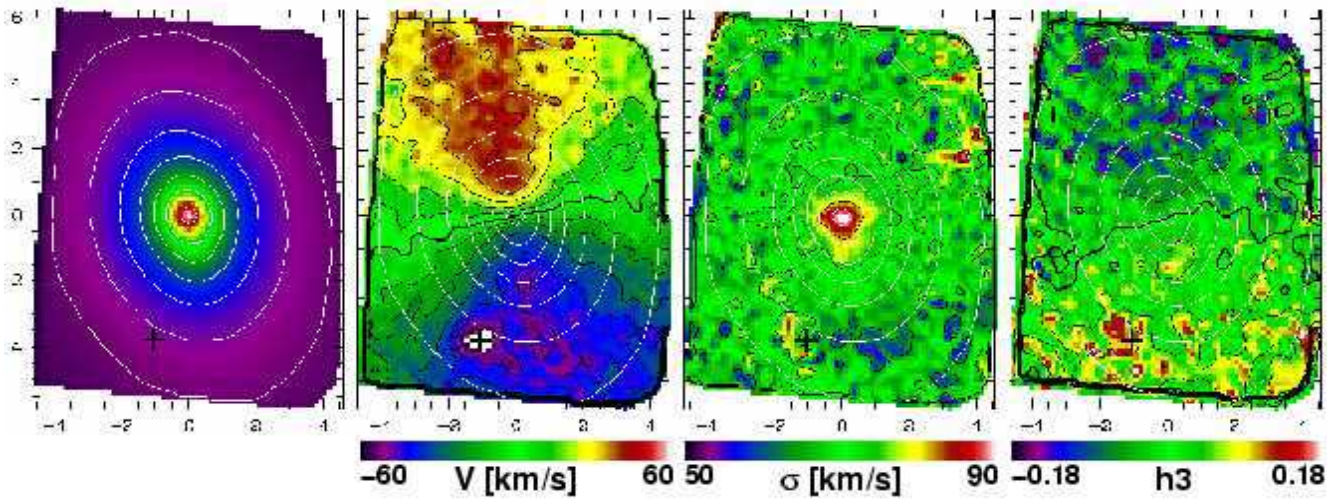,width=17.7truecm,clip=}}
\caption{\sauron\ measurements of M32, based on a single 2700~s
exposure with the HR mode. The field of view is $9''\times11''$, and
the spatial sampling is $0\farcs27\times0\farcs27$. From left to
right: a) reconstructed total intensity, with isophotal contours shown
in steps of 0.25 mag arcsec$^{-2}$.  b) Stellar mean velocity $V$,
with the zero-velocity contour indicated by the solid curve.  c)
stellar velocity dispersion $\sigma$.  d) Gauss--Hermite moment
$h_3$. The $+$ indicates the location of the planetary nebula
discussed in the text. }
\label{fig:m32-obs}
\end{figure*}

Many of the galaxies observed so far with \sauron\ show similar
evidence for a central stellar disk. However, most appear thicker than
in the case of NGC~3623. This is most likely due to projection
effects, the disks being more nearly face-on, but could also be due to
the presence of intrinsically thick disks. The maps of NGC 3623
provide a further illustration of the advantages of integral-field
spectroscopy. By contrast to spectroscopy along a few slits, the
\sauron\ maps allow us to delineate the entire (projected) kinematics
of any embedded subsystem.

\subsection{HR mode observations of M32}
\label{sec:m32}

M32 is a high-surface brightness, compact E3 companion of the
Andromeda galaxy. In the past two decades, M32 has been scrutinized
extensively for the presence of a central black hole, with ever-higher
spatial resolution and increasingly sophisticated dynamical modeling
(Dressler 1984; Tonry 1984, 1987; Dressler \& Richstone 1988; van der
Marel et al.\ 1994; Qian et al.\ 1995; Bender, Kormendy \& Dehnen
1996; van der Marel et al.\ 1997a, b, 1998; Joseph et al.\ 2001).

We observed M32 on October 15, 1999, with the principal aim of testing
the HR mode of \sauron. The seeing was $\approx$$0\farcs95$.
Figure~\ref{fig:m32-obs} shows the integrated intensity, stellar mean
velocity, velocity dispersion and Gauss-Hermite moment $h_3$ of the
central $9''\times 11''$, derived from a single 2700~s exposure. The
velocity field is very regular, has a peak amplitude of about 50~\kms,
and is consistent with axisymmetry. The $h_3$-field is similarly very
regular, and $h_3$ has the opposite sign of $V$ everywhere.  The
velocity dispersion in M32 is smaller than the HR-mode instrumental
dispersion of $\approx$105 \kms\ (Paper I), except in the inner
arcsecond (e.g., van der Marel et al.\ 1994, 1997a; Joseph et al.\
2001). For this reason it is difficult to measure $h_4$. It varies
little over the field, and we do not show it here.

\begin{figure}
\centerline{\psfig{file=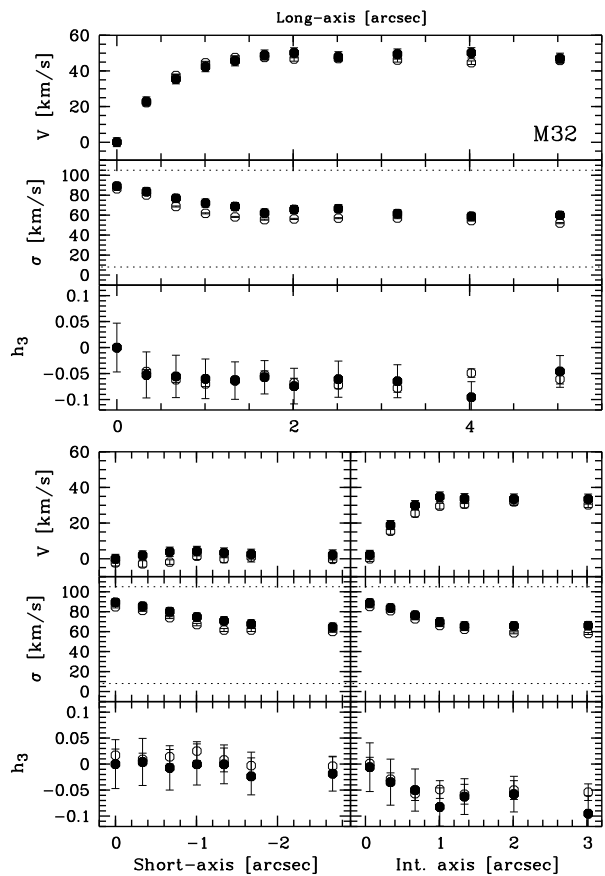,width=7.9truecm}}
\caption{Comparison of the \sauron\ kinematics of M32 (solid dots)
with the \isis\ long-slit measurements of van der Marel et al.\
(1994), taken with similar seeing but with superior spectral
resolution (open circles). Top: major axis, PA=160$^\circ$.  Bottom
left: minor axis, PA=71$^\circ$. Bottom right: intermediate axis
(PA=116$^\circ$). The data have been folded about the centre.  The
horizontal dotted lines in the $\sigma$-panels indicate the
instrumental resolutions. }
\label{fig:m32-comparison}
\end{figure}

The plus sign in the M32 maps indicates the position of strong [OIII]
emission, with a velocity close to the systemic velocity of M32. It is
not spatially resolved and most likely originates in a planetary
nebula in M32.

Figure~\ref{fig:m32-comparison} compares the \sauron\ measurements
with the long-slit observations of van der Marel et al.\ (1994), using
\isis\ at the same Cassegrain port, with similar seeing, but with an
instrumental dispersion of 8 km/s. The \sauron\ HR measurements were
axisymmetrized and binned in the same way as the \isis\ data. The
agreement is excellent: the differences \sauron\ -- \isis\ are
$2.3\pm2.2$ \kms\ in $V$, $6.0\pm 2.6$ \kms\ in $\sigma$, and
$-0.009\pm 0.016$ in $h_3$.  The observed $\sigma$ is well below the
\sauron\ instrumental dispersion, but can still be determined reliably
over the entire HR field of view, albeit with a small systematic
offset.

\begin{figure*}
\centerline{\psfig{file=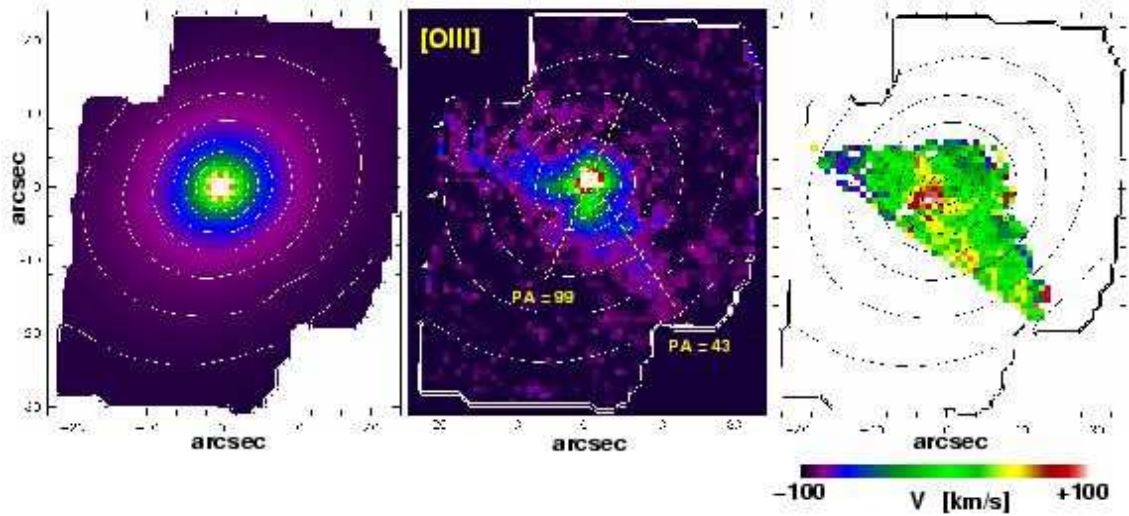,width=15.0cm,clip=}}
\caption{\sauron\ measurements of the [OIII] emission in NGC 5813.
From left to right: a) reconstructed total intensity of the stellar
continuum light. b) intensity of [OIII] emission. c) velocity field of
the emission-line gas.  The maps are based on two pointings of
$4\times 1800$~s each, sampled with $0\farcs8 \times 0\farcs8$ pixels.
The dashed lines in the middle panel indicate the position angles of
the H$\alpha$+N[II] long-slit spectroscopy by Caon et al.\ (2000),
shown in Figure~\ref{fig:n5813-gas-comp}. }
\label{fig:n5813-gas-maps}
\end{figure*}

Van der Marel et al.\ (1998) constructed fully anisotropic
axisymmetric models for M32 by means of Schwarzschild's (1979) orbit
superposition technique. The models fit the surface brightness
distribution, high signal-to noise groundbased long-slit kinematics
along multiple position angles, and \fos\ measurements. They require a
central black hole of mass $(3.4\pm0.7)\times 10^6 M_\odot$ but do not
constrain the inclination of M32 strongly: models with $i=55^\circ$
and $i=90^\circ$ fit the measurements equally well.

Our HR mode data shows that the stellar velocity field of M32 is
accurately symmetric with respect to the minor axis of the surface
brightness distribution, as expected in axisymmetric geometry for any
inclination angle. The position angle of the zero velocity curve is
within 2$^\circ$ of the minor axis of M32.  This suggests that there
is no compelling reason to consider triaxial models for M32, and that
the black hole mass determination is robust.

\section{Gas kinematics}
\label{sec:gas-kin}

The wavelength range 4810--5350\AA\ of the {\tt SAURON} spectra allows
observation of the H$\beta$, [NI] and [OIII] emission lines in nearby
objects. Paper I showed an example of the non-axisymmetric
distribution of [OIII] in NGC~3377. Here, we consider NGC~5813 and
compare the properties of the emission-line gas as derived from our
{\tt SAURON} observations with long-slit data from the literature. We
also discuss NGC~7742, in which we have detected a gaseous ring which
counter-rotates with respect to the central stellar component. We
first describe the algorithm we use to measure the emission
lines. This expands on the brief outline presented in \S5.3 of Paper
I.

\subsection{Measurement of the emission lines}

The emission lines need to be disentangled from the stellar absorption
line spectrum of a galaxy. We use an algorithm similar to the one
employed by Emsellem \& Goudfrooij (2001, in prep., see also
Goudfrooij \& Emsellem 1996), which consists of the following
iterative steps:

\begin{enumerate}
\itemsep 3pt
\item We first derive the stellar kinematics ($V$ and $\sigma$) using
the FCQ algorithm and a single stellar template spectrum (typically
that of a K0 giant), and truncate the spectral domain when the
brightest emission lines (e.g., [OIII], H$\beta$) influence the results.
The resulting maps are then median filtered to remove residual
spurious values due to, e.g., a low signal-to-noise ratio or, more
likely, the presence of a strong emission line.

\item We then build a spectral library, including both stellar and
galactic spectra (devoid of detectable emission lines) in the same
spectral domain but at higher spectral resolution than that of the
\sauron\ datacube. A linear combination of these spectra is fitted to
each spectrum in the \sauron\ datacube while masking out
regions where emission lines may contribute, and taking into account
the measured differences in $V$ and $\sigma$.

\item We then subtract the fitted spectra to obtain---in principle---a
pure emission-line datacube. All lines are simultaneously fitted using
Gaussians of the same velocity and width, by means of dedicated
software.

\item Finally, we subtract the fitted emission-line datacube from the
original \sauron\ datacube to obtain pure stellar absorption-line
spectra.
\end{enumerate}

This procedure generally works well with only two iterations. It is
very effective in recovering faint emission and superior to
narrow-band imaging.  Problems may occur when, for example, the
emission lines have non-Gaussian profiles, or when they do not probe
the same physical regions. These problems can be solved during
step~(iii) by generalizing the fitting functions and/or relaxing the
fitting constraints. A more serious problem may come from the lack of
proper stellar types in the spectral library (e.g. super metal-rich
stars). This can result in over- or underestimated emission-line
fluxes, particularly in the case of the faint [NI]$\lambda$5200
doublet usually lying close to the strong Mg{\it b} absorption
lines. As a result, the library has to be tuned to each galaxy and
sometimes even to different regions within a galaxy.

\subsection{The emission-line gas in NGC 5813}
\label{sec:n5813-gas}

NGC~5813 is a large and bright elliptical galaxy classified as E1-2 in
the RC3 ($M_B=-20.99$, $T=-4.5$ in LEDA). It is located in the
Virgo-Libra Cloud (Tully 1988), near NGC~5814 and NGC~5806.  The
surface brightness profile of NGC~5813 has a core with nuclear cusp
slope $\gamma^V_{\rm phys}=0.24\pm0.03$ (Carollo et al.\ 1997).  The
ellipticity increases with radius, and the galaxy displays isophotal
twists (Lauer et al.\ 1995; Carollo et al.\ 1997). NGC~5813 contains a
rapidly rotating central component within the inner $\lta7\arcsec$,
has negligible rotation at larger radii, and shows weak evidence for
minor-axis rotation (Efstathiou, Ellis \& Carter 1980, 1982). The
galaxy is undetected in HI or CO, but has an unresolved, weak central
radio continuum source (6~cm; Birkinshaw \& Davies 1985; Wrobel \&
Heeschen 1991). It was also detected by IRAS in the 12 and 100$\mu$m
bands (Thronson, Bally \& Hacking 1989), although many authors report
non-detections (e.g.\ Knapp et al.\ 1989). Emission-line ratios are
typical of LINERS but no X-ray emission is detected (Ho, Filippenko,
\& Sargent 1997).\looseness=-2

HST imaging of NGC 5813 has revealed a dust lane parallel to the major
axis, east of the nucleus, and dust filaments inside $7''$, i.e.,
inside the decoupled core (e.g., Carollo et al.\ 1997). The
distribution of emission-line gas, as traced by H$\alpha$+[NII]
narrow-band imaging, is very elongated and bent: it extends along the
minor axis at PA $=43^\circ$ on one side, but follows PA $=190^\circ$
on the other side (Caon, Macchetto \& Pastoriza 2000).  This is
similar to the shape of the region where the $B-R$ color is affected
by dust (Peletier 1989).\looseness=-2

We observed NGC~5813 on 29 \& 30 March and 4 April, 2000, obtaining a
mosaic of two fields exposed for $4\times1800$~s each, with overlap on
the nucleus. The seeing was $1''$ on the first two nights and
$2\farcs5$ on the last.  Figure~\ref{fig:n5813-gas-maps}{\it a} shows
the reconstructed stellar continuum intensity derived from our
\sauron\ mosaic.  Figure~\ref{fig:n5813-gas-maps}{\it b} shows the
integrated intensity map of [OIII]. Its morphology resembles that of
the H$\alpha$+[NII] map of Caon et al.\ (2000) and exhibits a complex
filamentary structure, most likely not (yet) in equilibrium with the
potential of the central region.  This is confirmed by the \sauron\
velocity map (Fig.~\ref{fig:n5813-gas-maps}{\it c}), which does not
display a simple pattern.

The \sauron\ gas kinematics can be compared with the kinematics of
H$\alpha$+N[II] obtained by Caon et al.\ (2000) along two position
angles. In order to be consistent with the \sauron\ kinematics (for
both the stars and the gas), we had to offset the Caon et al.\ data by
$+70$~\kms. The RMS deviations increase from $\approx17$ \kms\ in the
center to $\approx50$ \kms at the outermost measured points.

\begin{figure}
\centerline{\psfig{file=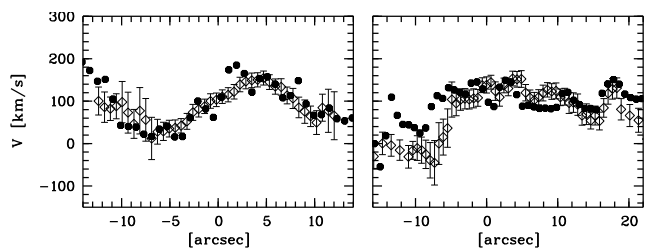,width=8.5cm,clip=}}
\caption{Comparison of the \sauron\ [OIII] velocities (filled squares)
with those derived by Caon et al.\ (2000), based on H$\alpha$+N[II]
measurements at PA=43$^\circ$ and PA=99$^\circ$ (open squares). We
have offset the Caon et al.\ velocity measurements by 70 \kms\
to be consistent with the \sauron\ kinematics. }
\label{fig:n5813-gas-comp}
\end{figure}

\begin{figure*}
\centerline{\psfig{file=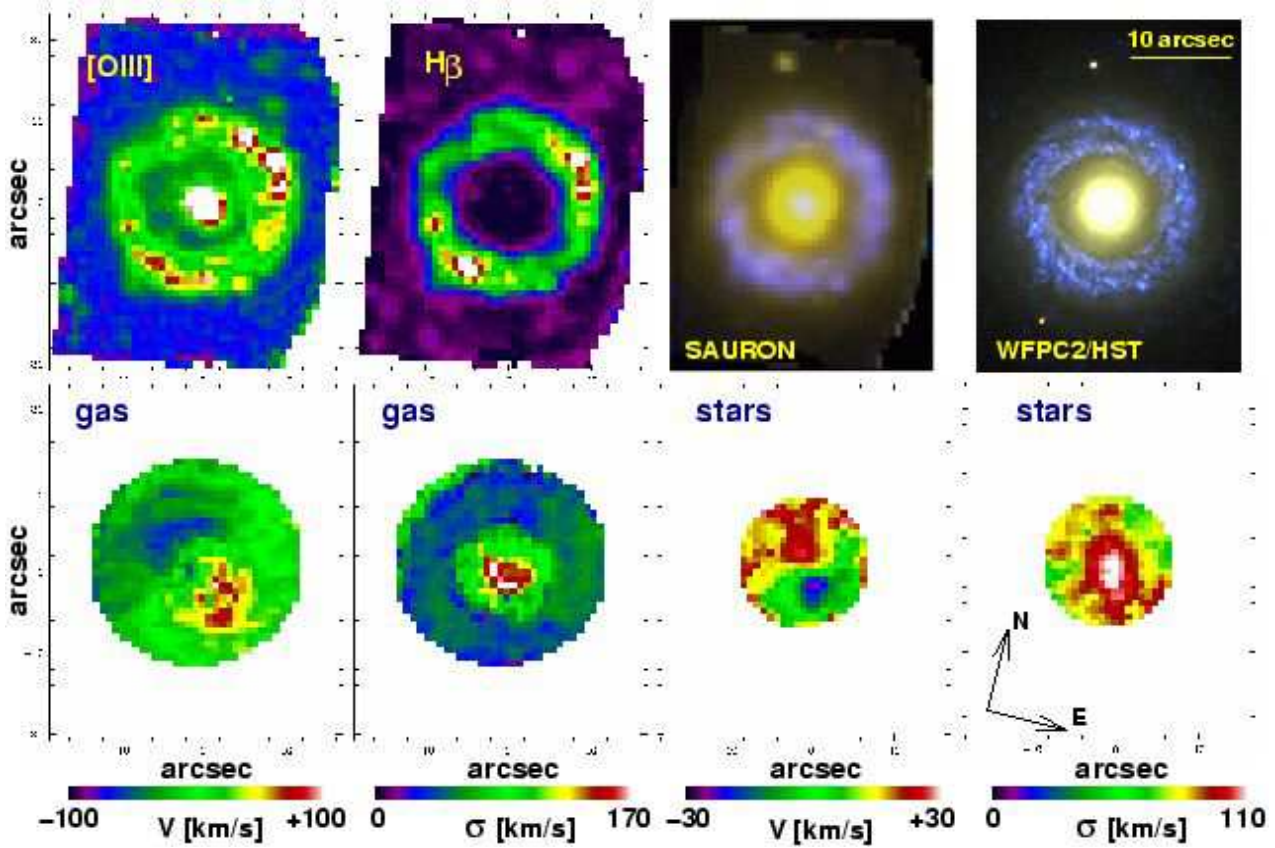,width=16.8cm,clip=}}
\caption{\sauron\ maps of NGC 7742, based on 1 pointing, exposed for
$3\times 1800 + 1\times 900$~s. The field-of-view is $33''\times 41''$
and the spatial sampling is $0\farcs8\times0\farcs8$. The top panels
show the emission-line intensity distributions of O[III] and H$\beta$,
followed by a colour-coded reconstructed image composed of [OIII]
(blue), blue continuum (green), red continuum (red) derived from the
\sauron\ data, and a similar colour-coded image composed of HST/WFPC2
exposures with the F336W (blue), F555W (green) and F814 (red) filters.
The bottom row shows (from left to right) the derived gas velocity and
velocity dispersion fields, and the stellar velocity and velocity
dispersion fields.\looseness=-2 }
\label{fig:n7742-obs}
\end{figure*}

\subsection{The extended emission-line gas in NGC 7742}
\label{sec:n7742}

NGC~7742 is a face-on spiral classified as S(r)b in the RC3
($M_B=-19.76$, $T=1.3$ in LEDA). It probably forms a binary system
with NGC~7743 in the Pegasus Cloud (Tully 1988; Honma 1999) and is a
good example of the latest spirals included in our sample. De
Vaucouleurs \& Buta (1980) first identified the inner stellar ring and
Pogge \& Eskridge (1993) detected a corresponding small, bright ring
of H~II regions with faint floculent spiral arms (see also Wozniak et
al.\ 1995). NGC~7742 also possesses a significant amount of neutral
hydrogen, molecular gas, and dust (e.g., Roberts et al.\ 1991).  It
shows essentially no sign of departure from axisymmetry (Rix \&
Zaritsky 1995; Kornreich, Haynes, \& Lovelace 1998). It is classified
as a transition LINER/H~II nucleus object (Ho, Filippenko, \& Sargent
1997).

We observed NGC~7742 with \sauron\ on the night of October 13,
1999. We obtained $3\times1800 + 1\times900$~s on one field centred on
the galaxy. The seeing was 1.5--2.5$''$. The spectra reveal emission
from H$\beta$, [OIII], and [NI]. The detection of the [NI] doublet is
rather difficult because we expect the line to be about ten times less
luminous than H$\beta$, and it lies close to the red edge of the Mg$b$
triplet.  Figure~\ref{fig:n7742-obs} shows the intensity in [OIII] and
H$\beta$, together with the derived velocity and velocity dispersion
fields. Most of the emission is confined to a ring which coincides
with the flocculent spiral arms. The H$\beta$ line is clearly dominant
in the ring, with [OIII]/H$\beta$ ranging from 0.06 to 0.14. We also
detect all three sets of lines at the centre, but this time with
[OIII]/H$\beta$ $\approx 1.0 \pm 0.1$.\looseness=-2

Also shown is a colour-coded reconstructed image composed of [OIII]
(blue), blue continuum (green), red continuum (red) derived from the
\sauron\ data cube, and a similar colour-coded image composed of
HST/WFPC2 exposures with the F336W (blue), F555W (green) and F814W
(red) filters (GO 6276, PI J.\ Westphal). The \sauron\ map does not
have the spatial resolution of HST, but it does demonstrate that our
analysis technique is capable of providing accurate emission-line
maps.

The main surprise comes from the comparison of the stellar and gas
kinematics: the gas in the ring counter-rotates with respect to the
stellar component inside the radius of the ring. The amplitude of the
stellar velocities is modest, because the galaxy is seen close to
face-on, but the zero velocity curve is well-defined, and oriented at
PA$=42^\circ\pm12^\circ$. This is consistent with the zero velocity
curve of the gas, which is at PA$=35^\circ\pm5^\circ$. The stellar
velocities outside and in the ring itself are consistent with being
equal to the systemic velocity, although any velocity difference of
the order of 30 \kms\ is difficult to detect in these regions due to
the lower signal-to-noise of the spectra and/or the emission line
contamination.  The signs of stellar and gas velocities inside the
ring are however clearly opposite. The ring-like structure in the gas
appears as a tightly wound multi-arm spiral structure, reminiscent of
nuclear ($< 1$~kpc) resonant rings (Buta \& Combes 1996).  High
resolution infrared images would be invaluable to attempt the
detection of a non-axisymmetric (weak bar) component in the older
stellar population. The spirals can be followed in the HST images into
the central arcsecond via the dust absorption features. If we assume
they correspond to a trailing wave, then the near side is in the
North-East quadrant.

\section{Line-strength maps}
\label{sec:line-strengths}

The \sauron\ wavelength coverage allows measurement of the
line-strength indices H$\beta$, Mg\,{\it b}, and Fe5270 in the
Lick/IDS system (Worthey et al.\ 1994; see also Paper~I). These
indices, in combination with simple stellar population models, can be
used to estimate luminosity-weighted metallicities and ages.
Two-dimensional line-strength maps make it possible to study the shape
of the contours of constant line-strength, and consequently of
constant metallicity and age (see, e.g., Peletier et al.\ 1999; del
Burgo et al.\ 2001). Davies et al.\ (2001) presented and discussed
such maps for NGC~4365. Here, we consider the observed line-strength
distributions in NGC~3384 and NGC~5813, based on the observations
described in \S\S\ref{sec:n3384-kin} and \ref{sec:n5813-gas}, and
compare them with previous long-slit measurements.

\begin{figure*}
\centerline{\psfig{file=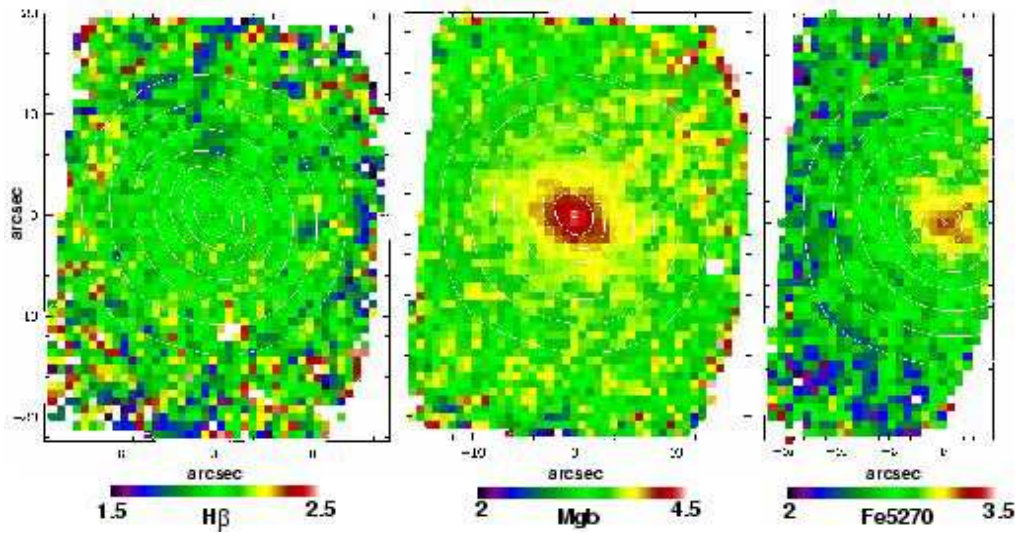,width=13.5cm,clip=}}
\caption{\sauron\ line-strength maps of NGC 3384, derived from the
exposures described in Figure~\ref{fig:n3384-obs-kin}. From left to
right: a) H$\beta$. b) Mg~{\it b}. c) Fe5270. Contours of constant
reconstructed intensity of the stellar continuum light are
superimposed. }
\label{fig:n3384-lines}
\end{figure*}

\subsection{Measurement of absorption line-strengths}

Line-strengths, as measured in the Lick/IDS system, are a measure of
the flux in a central bandpass compared to a continuum defined by two
sidebands. Two steps are required to convert measured equivalent
widths to line-strength indices on the Lick/IDS system: (i) a
correction for the difference in spectral resolution, and (ii) a
correction for the stellar velocity dispersion of the galaxies. We
correct for the difference in resolution by smoothing the data to the
resolution of the Lick/IDS system, 8.4 \AA\ in the \sauron\ wavelength
range (Worthey \& Ottaviani 1997). Absorption-line indices for stars
measured in this way should be on the Lick/IDS system, apart from
small offsets ($\lta 0.15$~\AA) that arise because the Lick/IDS system
is not based on flux-calibrated spectra (e.g., Vazdekis et al.\
1997). In future papers we will use our full sample of galaxy
observations to establish the significance of any offsets in the
line-strength indices compared to the Lick/IDS system.

In order to correct the line strengths for the broadening caused by
the stellar velocity distribution we convolved a template star
spectrum with the observed LOSVD of the galaxy at each position (see
\S\ref{sec:stellar-kin}), and derived the correction necessary to
evaluate the indices at zero velocity dispersion. This correction does
not depend critically on the stellar type or temperature for G and K
stars (e.g., Davies et al.\ 1993).

\subsection{The effect of sky subtraction errors}

Accurate sky subtraction is achieved using the 146 lenslets located
$1\farcm9$ away from the main field that record sky spectra
simultaneously with the object spectra (Paper I). Some of the galaxies
are so large that even at this separation the galaxy contribution is
non-negligible. The error caused by this `galaxy-contaminated' sky is
however smaller than the typical Poisson errors. For example, for
NGC~3384 the galaxy contribution in the sky field is roughly a factor
of 40 smaller than in the corner of the \sauron\ field
(25\arcsec). When one subtracts the spectrum in the sky field from a
spectrum at 25\arcsec, one also subtracts the galaxy contribution. In
the worst case this would introduce a relative line-strength error of
2.5\%.  Since, however, line-strength gradients are fairly gentle,
this is a generous overestimate of the error. For the larger galaxy
NGC~5813, the difference in galaxy surface brightness between the
edge of the \sauron\ field and the sky is only a factor of 10, which
means that the maximum relative error in the indices at the edge of
the frame is at most 10\%. In practice it will be smaller than 5\%.

\subsection{NGC 3384}
\label{sec:n3384-lines}

Figure~\ref{fig:n3384-lines} shows the maps of H$\beta$, Mg\,{\it b}\/
and Fe5270 for NGC~3384, obtained with the procedure described in the
above. Due to the effect of a slightly tilted interference filter, the
observed wavelength range varies over the field of view (Paper~I,
\S~4.9).  This variation affects only the extent of the spectra, and
impacts our measurements of the Fe5270 index for galaxies with
significant recession velocities. For most galaxies in our sample, the
wavelength range of the red sideband is included in the \sauron\/ data
for about 2/3 of the field of view. We illustrate this effect in
Figure~\ref{fig:n3384-lines}c. We have taken a conservative approach
showing only measurements which are completely unaffected by the
decrease in wavelength-coverage. In future work we intend to improve
the data-reduction procedure, and increase the effective field of view
for this index.

\begin{figure}
\centerline{\psfig{file=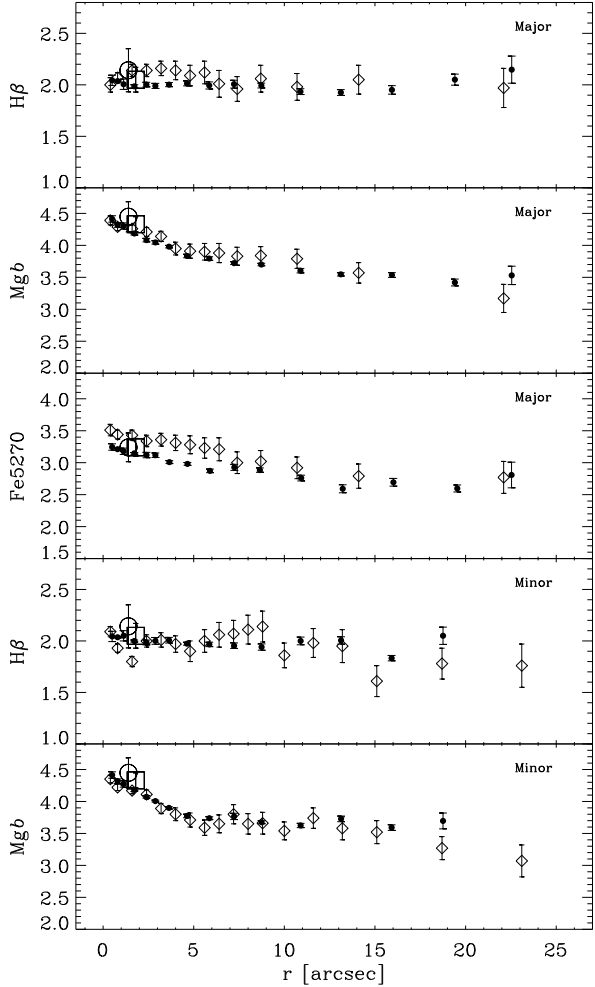,width=8.5cm,clip=}}
\caption{Comparison of the \sauron\ H$\beta$, Mg\,{\it b}\/ and Fe5270
line-strengths for NGC~3384 (dots) with the measurements by Fisher,
Franx \& Illingworth (1996, open diamonds) along the major (PA
$=53^\circ$) and the minor (PA $=143^\circ$) axis.  The large open
circle at $r=1\farcs4$ indicates the Lick observation (Trager et al.\
1998) and the large open square at $r=1\farcs8$ represents the data of
Kuntschner et al.\ (2001). The \sauron\/ data are binned
logarithmically in radius along a simulated slit of 2\arcsec\/ width.}
\label{fig:n3384-lines-comp}
\end{figure}

Fisher, Franx \& Illingworth (1996) published accurate line-strength
indices for NGC~3384, tabulating amongst other indices H$\beta$,
Mg\,{\it b}\/ and Fe5270 on the Lick/IDS system. Their spectra had
3.1~\AA\/ (FWHM) spectral resolution ($\sigma_{\rm inst}\simeq75$
\kms), and were taken with a 2\arcsec\/ slit along the major and minor
axes (PA $=53^\circ$ and $143^\circ$). Measurements of the Fe5270
index are only available for the major axis.
Figure~\ref{fig:n3384-lines-comp} shows the comparison with simulated
long-slit measurements derived from the {\tt SAURON} data-cubes. The
\sauron\/ data are binned logarithmically in radius to produce a
similar number of data points. Fisher et al.\ (1996) folded their
measurements about the centre and we have followed the same procedure
to make this comparison.

The agreement for both Mg\,{\it b}\/ and H$\beta$ is excellent.
Re-sampling our data to the spatial sampling of the Fisher et al.
(1996) data results in RMS deviations of 0.1~\AA\/ in Mg\,{\it b}\/
and 0.1~\AA\/ in H$\beta$. The systematic offsets are less than
0.03~\AA.  The comparison for the Fe5270 index shows a systematic
offset of $\sim$0.24~\AA\/ in the sense that our data show weaker
Fe5270, which is however more in line with the expected value for a
galaxy of this luminosity (Davies et al.\ 1993). The RMS deviations
are 0.1~\AA. The offset is modest and the overall gradient is
consistent with the data of Fisher et al.\ (1996). There is good
agreement with the central measurements by Trager et al.\ (1998, large
open circle) and Kuntschner et al.\ (2001, large open square).  The
agreement between the Fe5270 index in NGC~5813 (see below) and
published data re-inforce our confidence in these measurements.

\begin{figure}
\centerline{\psfig{file=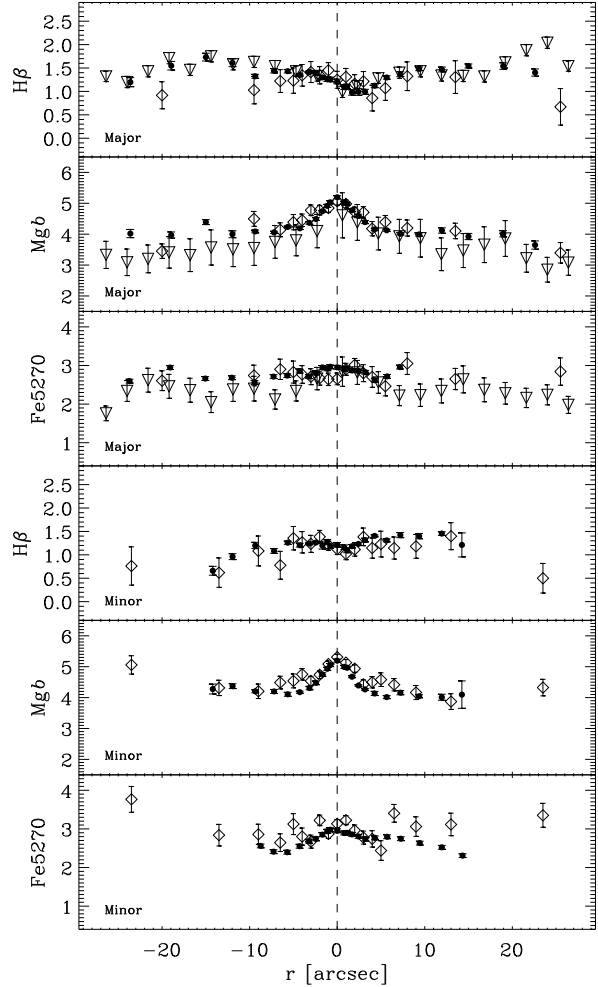,width=8.5cm,clip=}}
\caption{Comparison of the \sauron\ H$\beta$, Mg\,{\it b}\/ and Fe5270
line-strengths for NGC~5813 (3\arcsec\/ simulated slit, filled
circles) with the measurements by Gorgas et al.\ (1990, Run~2, open
triangles) and Gonz\'alez (1993, open diamonds), for the major and
minor axis.}
\label{fig:n5813-lines-comp}
\end{figure}

\begin{figure*}
\centerline{\psfig{file=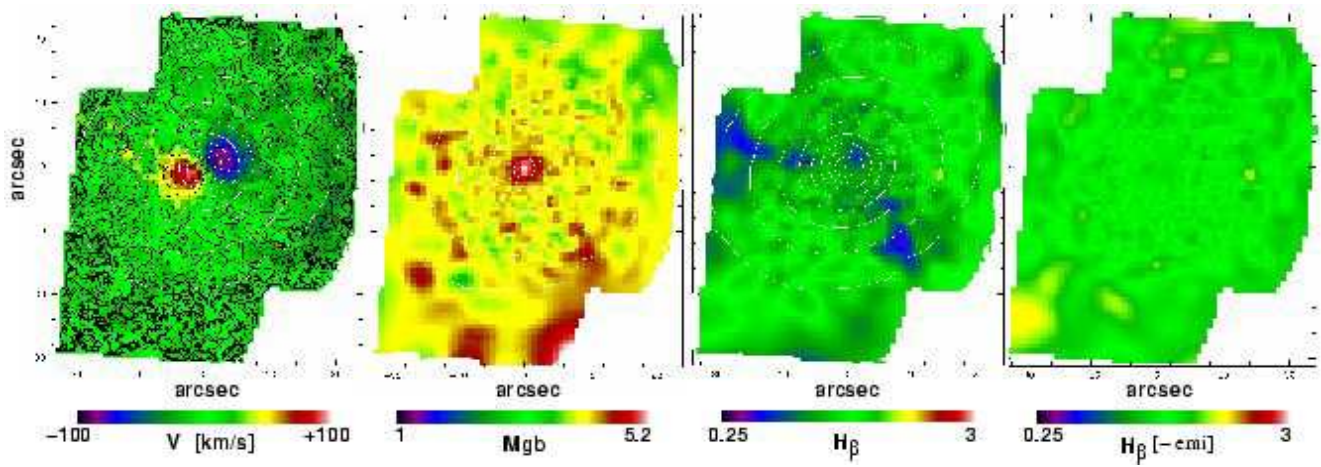,width=17.5cm,clip=}}
\caption{\sauron\ maps of NGC 5813, derived from the two pointings
described in Figure~\ref{fig:n5813-gas-comp}. From left to right: a)
stellar absorption-line velocity field. b) Mg~{\it b}.  c) H$\beta$
observed. d) H$\beta$ corrected for emission.  The contours of
constant integrated intensity of the continuum light are
superimposed. }
\label{fig:n5813-lines}
\end{figure*}

The {\tt SAURON} maps show that Mg\,{\it b}\/ is enhanced in the
region defined by the inner major-axis disk ($r\le 5$\arcsec). Outside
this region there is a weaker gradient. This supports the
interpretation that later infalling components (i.e., the material
that formed the disk) are made out of enriched material (as in
NGC~4365, Davies et al.\ 2001). The same is seen in Fisher et al.'s
data. In the region between 4\arcsec\/ and 10\arcsec, the contours of
equal Mg\,{\it b}\/ line-strengths have the same shape as the
isophotes of the stellar surface brightness distribution. H$\beta$ is
roughly constant everywhere (as in NGC~4365). However, NGC~3384
shows an unusually strong H$\beta$ index, $\sim$2.0~\AA\/, which
indicates a luminosity-weighted age of $\sim$5~Gyr younger than that
of the oldest elliptical galaxies. Unlike the disk, the other
prominent central component, the inner bar (e.g., Busarello et al.\
1996) is not seen in the line-strength maps. Martin \& Roy (1994)
proposed that bars smear out population gradients, based on a
correlation between bar strength and radial O/H abundance gradient in
the young stellar population derived from individual H{\sc II}
regions. To date, it has not been possible to verify this correlation
using gradients in the absorption line-strengths, which sample the
entire stellar population. Statistical analysis for the entire
\sauron\ sample will have to show whether radial abundance gradients
are affected by the presence of bars.

\subsection{NGC~5813}
\label{sec:n5813-lines}

Early measurements of the line-strength profile in NGC~5813 are
reported in Efstathiou \& Gorgas (1985), and modeled in Arag\'on,
Gorgas \& Rego (1987). Superior measurements were published by Gorgas,
Efstathiou \& Aragon--Salamanca (1990), based on a 34200 sec major
axis exposure on the AAT with the {\tt IPCS} and a 3\arcsec\/ slit
(cf.\ Efstathiou et al.\ 1980; 1982).  They present, among other
indices, the H$\beta$, Mg\,{\it b}\/ and Fe5270 index.  Gonz\'alez
(1993) also reports these indices, for the major and minor axis
(2\farcs1 slit).

Figure~\ref{fig:n5813-lines-comp} shows the comparison of the \sauron\
measurements with those by Gorgas et al.\ (1990) and Gonz\'alez
(1993). The overall agreement is good. The comparison for H$\beta$ is
excellent, even in the central regions where nebular emission severely
affects the H$\beta$ index. This can be seen at $r \approx 2\farcs5$
where all datasets show a dip in H$\beta$ absorption
strength. Re-sampling our data to the spatial sampling of the
published data results in RMS deviations of $\sim$0.25~\AA\/ and
systematic offsets $\le 0.1$~\AA\/ for the Gonz\'alez data. The
comparison with Gorgas et al.\ shows RMS deviations of up to
0.45~\AA\/ and systematic offsets of up to 0.6~\AA.  However, this is
consistent with the error bars given by Gorgas et al. Overall the
comparison for NGC~5813 shows larger RMS deviations than the
comparison for NGC~3384. We ascribe this to the effects of nebular
emission and slightly larger errors in the literature data.

Figure~\ref{fig:n5813-lines} displays the {\tt SAURON} maps of mean
stellar velocity, Mg\,{\it b}, H$\beta$ observed and H$\beta$
corrected for emission for NGC~5813, in all cases with contours of the
stellar continuum intensity superimposed. The rapidly rotating core is
clearly visible, and extends $\approx$12\arcsec\/ along the major axis
and $\approx$7\arcsec\/ along the minor axis. The zero-velocity curve
is nearly straight, and is misaligned by $13\pm2^\circ$ from the minor
axis of the surface brightness distribution. The Mg\,{\it b}\/
distribution has a peak in the region of the rotating central disk,
and a very shallow gradient outside it, similar to NGC~3384 and
NGC~4365. H$\beta$ here is severely affected by emission lines (see
\S\ref{sec:gas-kin}), so need to be corrected for the emission before
studying the stellar populations.  The two-dimensional structure of
the emission can clearly be seen in the right-hand panel of
Figure~\ref{fig:n5813-lines}. We have subtracted the emission
contamination from the spectra by means of an optimal template fitting
routine. The `emission-free' H$\beta$ map is shown in the right-most
panel of Figure~\ref{fig:n5813-lines}. This shows that the stellar
absorption of H$\beta$ is roughly constant over the whole field,
similar to what we found for NGC~4365 (Davies et al.\ 2001).

\section{Conclusions}
\label{sec:conclusions}

We built \sauron\ to measure the intrinsic shapes and internal
velocity and metallicity distributions of early-type galaxies. This is
a key component of a comprehensive study of the internal structure of
these systems, aimed at gaining insight into the connection between
the kinematics and stellar populations, and the relation between
nuclear and global properties. Accordingly, we are carrying out a
representative survey of nearby early-type galaxies covering a large
range of luminosity, ellipticity, and so covering a range
of environment, nuclear cusp slope, and rotational support.

The full survey will not be completed until 2002, but the first
results presented here demonstrate that the \sauron\ measurements
agree with previous high-quality long-slit kinematic and line-strength
measurements. This establishes the scientific integrity of the
instrument. It also shows that integral-field spectroscopy is a mature
technique, superior to long-slit studies because of the fundamental
advantage of complete spatial coverage.

A preliminary analysis of the survey maps shows that early-type
galaxies display a variety of line-strength distributions and
kinematic structure which appears richer than often assumed. We have
shown specific examples of minor axis rotation, decoupled cores,
central stellar disks, and non-axisymmetric and counter-rotating
gaseous disks.  The provisional indication is that only a small
fraction of these galaxies can have axisymmetric intrinsic shapes.
The line-strength distributions appear to follow the surface
brightness distribution closely.

We have started to complement the \sauron\ maps with
high-spatial-resolution spectroscopy of the nuclear regions, using
\oasis. \stis\ spectroscopy for many of the sample galaxies is
available in the HST archive. Radial velocities of planetary nebulae
and/or globular clusters in the outer regions have been obtained for
some of the galaxies in our sample, and many more will become
available to about 5$R_e$ with a special-purpose instrument now under
construction (Freeman et al.\ 2001, in prep).

We are analyzing the \sauron\ maps by Fourier techniques. This will
provide the ellipticity, position-angle, and deviations from the
elliptic shape of the contours of constant velocity dispersion, $h_4$,
and line-strengths, as a function of radius (Copin et al.\ 2001, in
prep.). This approach has been customary for the analysis of the
surface brightness distribution of early-type galaxies (e.g, Kormendy
\& Djorgovski 1989). The maps of mean velocity and $h_3$ can be
quantified in a similar way, just as is possible for HI and CO data
(Franx, van Gorkom \& de Zeeuw 1994; Schoenmakers, Franx \& de Zeeuw
1997; Wong 2000; see also Copin 2000, Copin et al.\ 2001, in
prep.). Doing this for the entire sample will provide a statistically
reliable determination of the incidence of kinematically decoupled
cores, embedded disks, and minor axis rotation, and will also provide
parametrized input for statistical studies of, e.g., the distribution
of intrinsic shapes (Franx, Illingworth \& de Zeeuw 1991; Bak \&
Statler 2000) and the construction of dynamical models.

The detailed measurements for individual objects will be compared with
fully general galaxy models constructed by means of Schwarzschild's
(1979) numerical orbit superposition method (cf.\ Rix et al.\ 1997;
van der Marel et al.\ 1998; Cretton et al.\ 1999).  The modeling uses
all appropriate imaging and spectral data available, including HST and
\oasis\ spectra, to constrain the mass of a central black hole and the
orbital structure in the main body of the galaxy.  When combined with
the constraints on the stellar populations derived from the
line-strength distributions (Kuntschner \& Davies 1998), this will
shed new light on the fundamental connections between the large and
small scale dynamics, the formation (and existence) of supermassive
BHs and galactic nuclei, and the history of metal enrichment in
early-type galaxies.

\thanks It is a pleasure to thank the ING staff, in particular Rene
Rutten, Tom Gregory and Chris Benn, for enthusiastic and competent
support on La Palma. This project made use of the LEDA database: {\tt
www-obs.univ-lyon1.fr}. This paper benefitted from logistic support by
Kirsten Kol-Groen and Ruud Witmer, and from a careful reading by
Michele Cappellari and Davor Krajnovic. BM acknowledges support from
the International Gemini Observatory. RLD gratefully acknowledges the
award of a Research Fellowship from the Leverhulme Trust. The \sauron\
project is made possible through grants 614.13.003 and 781.74.203 from
ASTRON/NWO and financial contributions from the Institut National des
Sciences de l'Univers, the Universit\'e Claude Bernard Lyon I, the
universities of Durham and Leiden, the British Council, PPARC grant
`Extragalactic Astronomy \& Cosmology at Durham 1998--2002', and the
Netherlands Research School for Astronomy NOVA.

{}

\appendix

\section{The sample}

\begin{table*}
\caption{Properties of the `cluster' galaxies in the representative sample.}
\label{tab:props-cluster}
\begin{tabular}{clrrcccccccrrcc}
\hline
\hline
Galaxy & Type & $T$ & $V_{\rm sys}$ & $\Delta m$ & $M_B$ & $\null
\!\!\!\!(B\!-\!V)_e \!\!\!$ & Mg$_2$ & $R_e$
      & $\epsilon_{25}$ & $\mu_e$ & $\null\!V_{\rm max}\!$ & $\sigma$
      & $\null\!\!\gamma^V_{\rm phys}\!\!$ & $M_{\rm BH}$  \\
(1) & ~~(2) & (3) & (4) & (5) & (6) & (7) & (8) & (9) & (10) & (11) &  (12) & (13) & (14) & (15)  \\
\hline
\multicolumn{5}{l}{Ellipticals}\\
\noalign{\smallskip}
NGC3377 &E5-6        &-4.0 & 698 &30.14 &-19.24 &0.905 &0.278 &34 &0.39 &20.98 & 89 &136 &1.15 &0.6-2.5(8)\\
NGC3379 &E1          &-4.0 & 877 &30.14 &-20.16 &0.975 &0.322 &35 &0.08 &20.72 & 49 &206 &     &0.5-1.6(8)\\
NGC4278 &E1-2        &-4.6 & 612 &30.68 &-19.93 &0.960 &0.306 &34 &0.06 &20.97 & 53 &252 &0.53 &  \\
NGC4374 &E1          &-3.5 &1016 &31.06 &-21.23 &1.000 &0.319 &51 &0.12 &20.68 & 17 &297 &     &0.9-2.8(9)\\
NGC4387 &E           &-3.4 & 550 &31.06 &-18.34 &0.965 &0.252 &16 &0.34 &      & 57 &117 &0.72 &  \\
NGC4458 &E0-1        &-3.8 & 676 &31.06 &-18.42 &0.915 &0.233 &26 &0.06 &      & 20 &102 &1.40 &  \\
NGC4473 &E5          &-4.2 &2210 &31.06 &-20.26 &0.990 &0.316 &26 &0.38 &20.18 & 54 &191 &     &  \\
NGC4486 &E0-1$^+$pec &-4.0 &1272 &31.06 &-21.79 &0.980 &0.303 &95 &0.30 &21.30 & 16 &351 &     &1.8-3.2(9)\\
NGC4552 &E0-1        &-3.4 & 288 &31.06 &-20.58 &1.000 &0.340 &29 &0.09 &20.25 &  9 &264 &0.53 &  \\
NGC4564 &E           &-4.1 &1116 &31.06 &-19.39 &0.965 &0.345 &20 &0.45 &20.43 &147 &168 &     &4.0-7.0(7)\\
NGC4621 &E5          &-4.0 & 431 &31.06 &-20.64 &0.975 &0.293 &40 &0.24 &20.72 &120 &245 &     &  \\
NGC4660 &E           &-4.1 &1082 &31.06 &-19.22 &0.990 &0.306 &12 &0.21 &19.76 &149 &191 &     &  \\
 & \\
\multicolumn{5}{l}{Lenticulars}\\
\noalign{\smallskip}
NGC3384 &SB0$^-$(s):         &-2.6 & 729 &30.14 &-19.56 &0.955 &0.306 &25 &0.49 &19.82 &    &142 &     &0.9-2.5(7)\\
NGC3489 &SAB0$^+$(rs)$\!\!$  &-2.1 & 688 &30.14 &-19.32 &0.845 &0.193 &20 &0.38 &19.60 &157 &138 &     &  \\
NGC4150 &S0$^0$(r)?          &-2.4 & 232 &30.68 &-18.48 &0.830 &0.131 &18 &0.30 &      &    &148 &     &  \\
NGC4262 &SB0$^-$(s)          &-2.6 &1361 &31.06 &-18.88 &0.970 &0.325 &13 &0.09 &20.03 &372 &186 &     &  \\
NGC4270 &S0                  &-1.1 &2349 &31.06 &-18.28 &0.950 &0.237 &13 &0.53 &20.79 &126 &140 &     &  \\
NGC4382 &S0$^+$(s)pec$\!$    &-1.8 & 745 &31.06 &-21.28 &0.895 &0.261 &55 &0.22 &20.74 &    &177 &     &  \\
NGC4459 &S0$^+$(r)           &-2.0 &1182 &31.06 &-19.99 &0.970 &0.270 &35 &0.23 &20.99 &    &174 &     &5.9-8.7(7)\\
NGC4477 &SB0(s):?            &-1.8 &1327 &31.06 &-19.96 &0.970 &      &   &0.09 &21.22 &    &172 &     &  \\
NGC4526 &SAB0$^0$(s):$\!$    &-1.6 & 566 &31.06 &-20.68 &0.975 &0.304 &44 &0.63 &20.68 &136 &256 &     &  \\
NGC4546 &SB0$^-$(s):         &-2.6 &1036 &31.06 &-19.98 &0.990 &      &27 &0.50 &20.36 &152 &242 &     &  \\
NGC4550 &SB0$^0$:sp          &-2.3 & 407 &31.06 &-18.83 &0.890 &0.191 &15 &0.71 &      &110 & 80 &     &  \\
NGC4570 &S0~sp               &-1.7 &1731 &31.06 &-19.54 &0.970 &0.342 &18 &0.68 &19.93 &    &188 &     &  \\
 & \\
\multicolumn{5}{l}{Spirals}\\
\noalign{\smallskip}
NGC3623 &SABa(rs)                  & 1.0 & 769 &30.14 &-20.82 &0.975 &     & 85 &0.76 &      & 234 &173 &     &  \\
NGC4235 &Sa(s)sp                   & 1.1 &2332 &31.06 &-19.20 &1.005 &     & 27 &0.78 &21.17 & 136 &    &     &  \\
NGC4245 &SB0/a(r):                 & 0.4 & 837 &30.68 &-18.72 &0.920 &     & 24 &0.17 &21.18 & 161 &    &     &  \\
NGC4274 &(R)SBab(r)$\!\!\!\!$      & 1.4 & 877 &30.68 &-20.08 &0.980 &     & 45 &0.63 &20.85 & 218 &138 &     &  \\
NGC4293 &(R)SB0/a(s)$\!\!\!\!\!\!$ & 1.4 & 823 &31.06 &-20.37 &0.940 &     & 83 &0.48 &22.67 & 163 &163 &     &  \\
NGC4314 &SBa(rs)                   & 1.5 & 957 &30.68 &-19.55 &0.895 &     & 34 &0.05 &21.11 & 204 &    &     &  \\
NGC4383 &Sa\$~pec                  & 1.1 &1663 &31.06 &-18.93 &0.455 &     & 11 &0.49 &19.82 &  96 &    &     &  \\
NGC4405 &S0/a(rs):                 & 1.3 &1751 &31.06 &-18.54 &0.750 &     &  - &0.34 &      &  68 &    &     &  \\
NGC4425 &SB0$^+$:sp                & 0.5 &1865 &31.06 &-18.83 &0.930 &     & 23 &0.64 &21.28 &     &132 &     &  \\
NGC4596 &SB0$^+$(r)                &-0.3 &1901 &31.06 &-19.94 &0.970 &     & 38 &0.13 &21.28 &     &152 &     &0.4-1.2(8)\\
NGC4698 &Sab(s)                    & 1.4 &     &31.06 &-20.05 &0.940 &     & 33 &0.31 &21.10 & 245 &172 &     &  \\
NGC4772 &Sa(s)                     & 1.2 &1042 &31.06 &-19.56 &0.930 &     & 24 &0.42 &20.30 & 239 &    &     &  \\
\hline
\end{tabular}
Notes:
(1)~Galaxy identifier.
(2)~Hubble type (NED).
(3)~Numerical morphological type (LEDA)
(4)~Recession velocity $V_{\rm sys}$ in \kms\ (LEDA).
(5)~Distance modulus $\Delta m$ in mag, determined as described in
    \S\ref{sec:properties}.
(6)~Absolute blue magnitude $M_B$ in mag.
(7)~Effective $(B-V)_e$ color in mag (LEDA).
(8)~Central Mg$_2$ linestrength in mag (Hypercat).
(9)~Effective radius $R_e$ in the $B$ band, in arcsec (RC3).
(10)~Ellipticity $\epsilon_{\rm 25}$ of the contour of 25 mag~arcsec$^{-2}$
     surface brightness (LEDA).
(11)~Mean effective surface brightness $\mu_e$ in the $B$ band, in mag (LEDA).
(12)~Maximum observed mean rotation $V_{\rm max}$ in \kms. For the
     ellipticals this is the maximum mean stellar rotation of the main galaxy
     (see \S\ref{sec:properties}). For the lenticulars  and the spirals
     the values taken from LEDA generally refer to the circular velocity.
(13)~Central velocity dispersion $\sigma$ in \kms (LEDA).
(14)~Average logarithmic slope $\gamma^V_{\rm phys}$ of the deprojected
     luminosity profile within 10--50 pc, taken from Carollo et al.\
     (1997, Tables 5 and 8).
(15)~Reported central black hole mass $M_{\odot}$, where the notation
     $a-b(c)$ indicates means $a-b\times10^c M_{\odot}$. References
    given in \S\ref{sec:properties}.
\end{table*}

Tables~\ref{tab:props-cluster} and \ref{tab:props-field} list the
galaxies in the representative sample defined in \S\ref{sec:sample},
for E/S0/Sa galaxies in `clusters' and in the `field',
respectively. The tables also list a number of global and nuclear
properties. In addition to the most common name used in LEDA, we give
the Hubble type (NED), numerical morphological type $T$ (LEDA),
recession velocity $V_{\rm sys}$ (LEDA), adopted distance modulus
$\Delta m$ (\S\ref{sec:sample}), absolute blue magnitude $M_B$
(\S\ref{sec:sample}), effective $B-V$ color $(B-V)_e$ (LEDA), central
Mg line-strength Mg$_2$ (Hypercat), effective radius $R_e$ in $B$
(RC3), ellipticity at $\mu_B=25$~mag~arcsec~$^{-2}$ $\epsilon_{25}$
(LEDA), mean $B$ effective surface brightness $\mu_e$ (LEDA), maximum
rotation velocity $V_{\rm max}$ (LEDA), central velocity dispersion
$\sigma$ (LEDA), deprojected central cusp slope $\gamma^V_{\rm phys}$
(\S\ref{sec:properties}), and central black hole mass $M_{\rm BH}$
(\S\ref{sec:properties}).

\begin{table*}
\caption{Properties of the `field' galaxies in the representative
         sample. See text for explanation.}
\label{tab:props-field}
\begin{tabular}{clrrccccrccrrcc}
\hline
\hline
Galaxy & Type & $T$ & $V_{\rm sys}$ & $\Delta m$ & $M_B$ & $\null\!\!\!\!(B\!-\!V)_e\!\!\!$ & Mg$_2$ & $R_e$
       & $\epsilon_{25}$ & $\mu_e$ & $\null\!V_{\rm max}\!$ &$\sigma$
       & $\null\!\!\gamma^V_{\rm phys}\!\!$ & $M_{\rm BH}$  \\
 (1) & ~~(2) & (3) & (4) & (5) & (6) & (7) & (8) & (9) & (10) & (11) &  (12) & (13) & (14) & (15)  \\
\hline
\multicolumn{5}{l}{Ellipticals}\\
\noalign{\smallskip}
NGC~821 &E6?    &-4.2 &1742 &31.86 &-20.44 &1.020 &0.316 &50 &0.32 &22.02 & 91 &208 &     &3.0-7.0(7)\\
NGC2699 &E:     &-5.0 &1825 &31.83 &-18.85 &0.980 &0.282 &   &0.06 &      &    &    &     &  \\
NGC2768 &E6:    &-3.1 &1324 &31.66 &-21.15 &0.960 &0.276 &64 &0.42 &21.94 &148 &188 &     &  \\
NGC2974 &E4     &-3.6 &1983 &31.93 &-20.32 &1.005 &0.305 &24 &0.39 &20.74 &207 &229 &     &  \\
NGC3608 &E2     &-4.3 &1201 &30.96 &-19.54 &1.000 &0.329 &34 &0.21 &21.57 & 26 &204 &0.80 &0.8-2.5(8)\\
NGC5198 &E1-2:  &-3.4 &2514 &32.80 &-20.38 &0.985 &0.322 &26 &0.14 &21.89 &  4 &195 &     &  \\
NGC5813 &E1-2   &-4.5 &1917 &32.10 &-20.99 &1.010 &0.319 &57 &0.24 &22.12 &  8 &238 &0.24 &  \\
NGC5831 &E3     &-4.3 &1660 &31.79 &-19.73 &0.985 &0.301 &26 &0.13 &21.48 & 27 &175 &     &  \\
NGC5838 &S0$^-$ &-3.0 &1348 &31.36 &-19.87 &1.010 &      &23 &0.59 &20.51 &225 &274 &     &  \\
NGC5845 &E:     &-4.1 &1581 &31.69 &-18.58 &1.120 &0.319 &   &0.32 &19.84 &127 &221 &0.51 &2.5-5.0(8)\\
NGC5846 &E0-1   &-4.2 &1722 &31.98 &-21.24 &1.030 &0.333 &63 &0.06 &21.96 &  7 &250 &     &  \\
NGC5982 &E3     &-3.9 &2876 &33.11 &-21.46 &0.940 &0.303 &24 &0.30 &21.08 & 85 &251 &0.21 &  \\
 & \\
\multicolumn{5}{l}{Lenticulars}\\
\noalign{\smallskip}
NGC~474 &S0$^0$(s)               &-2.2 &2304 &32.50 &-20.42 &0.940 &      &34 &0.19 &21.99 &244 &170 &     &  \\
NGC~524 &S0$^+$(rs)              &-1.5 &2422 &32.58 &-21.40 &1.070 &0.307 &50 &0.01 &21.73 &    &245 &0.52 &  \\
NGC1023 &SB0$^-$(rs)             &-2.6 & 607 &30.06 &-20.42 &1.010 &0.311 &40 &0.56 &20.32 &138 &206 &0.74 &2.4-4.8(7)\\
NGC2549 &S0$^0$(r)~sp            &-2.0 &1064 &31.12 &-19.36 &0.955 &0.269 &17 &0.68 &20.05 &    &146 &     &  \\
NGC3414 &S0~pec                  &-2.5 &1441 &31.52 &-19.78 &0.930 &0.327 &21 &0.17 &20.53 &223 &246 &     &  \\
NGC2685 &(R)SB0$^+$pec$\!\!\!\!$ &-0.7 & 867 &30.79 &-19.05 &0.935 &0.229 &32 &0.51 &21.50 &147 & 99 &     &  \\
NGC2695 &SAB0$^0$(s)             &-2.4 &1831 &31.83 &-19.38 &0.930 &0.319 &16 &0.27 &20.96 &    &220 &     &  \\
NGC3032 &SAB0$^0$(r)             &-1.7 &1561 &31.68 &-18.77 &0.630 &0.129 & 9 &0.11 &20.01 &169 & 82 &     &  \\
NGC3156 &S0:                     &-2.4 &1174 &30.90 &-18.08 &0.770 &0.110 &14 &0.38 &20.93 &362 & 84 &     &  \\
NGC5308 &S0$^-$~sp               &-1.1 &2029 &32.26 &-20.27 &0.925 &      &18 &0.82 &19.87 &146 &260 &     &  \\
NGC7332 &S0~pec~sp               &-1.7 &1197 &31.42 &-19.93 &0.905 &0.242 &15 &0.73 &19.43 &131 &134 &1.00 &  \\
NGC7457 &S0$^-$(rs)?             &-2.2 & 758 &30.46 &-18.81 &0.900 &0.178 &32 &0.41 &21.57 &    & 75 &     &1.7-6.0(6)\\
&  \\
\multicolumn{5}{l}{Spirals}\\
\noalign{\smallskip}
NGC1056 &Sa:                     & 1.0 &1695 &31.70 &-19.44 &1.020 &      &   &0.44 &      &140 &    &     &  \\
NGC2273 &SBa(r):                 & 0.5 &1887 &32.17 &-20.21 &0.940 &      &26 &0.25 &21.61 &228 &124 &     &  \\
NGC2844 &Sa(r):                  & 0.2 &1488 &31.66 &-18.38 &0.810 &      &   &0.55 &      &145 &114 &     &  \\
NGC4220 &S0$^+$(r)               &-0.1 & 966 &30.88 &-19.01 &0.970 &      &30 &0.61 &21.36 &167 &140 &     &  \\
NGC4369 &(R)Sa(rs)               & 1.1 & 986 &31.01 &-18.96 &0.500 &      &16 &0.01 &20.41 &288 &    &     &  \\
NGC5448 &(R)SABa(r)$\!\!\!$      & 1.2 &1971 &32.36 &-20.78 &0.760 &      &72 &0.48 &22.01 &207 &    &     &  \\
NGC5475 &Sa~sp                   & 1.0 &1699 &32.06 &-19.39 &      &      &   &0.74 &      &    &    &     &  \\
NGC5689 &SB0$^0$(s):             & 0.4 &2147 &32.53 &-20.32 &      &      &   &0.71 &      &170 &203 &     &  \\
NGC5953 &Sa:pec                  &-0.4 &2061 &32.22 &-19.61 &      &      &   &0.26 &      &120 &129 &     &  \\
NGC6501 &S0$^+$:                 &-0.3 &2958 &33.03 &-20.38 &1.050 &      &   &0.10 &      &193 &222 &     &  \\
NGC7731 &(R)SBa:pec$\!\!\!$      & 1.2 &2866 &32.96 &-18.78 &0.805 &      &15 &0.19 &21.43 &130 &    &     &  \\
NGC7742 &Sb(r)                   & 1.3 &1655 &31.82 &-19.76 &0.775 &      &15 &0.05 &20.23 & 92 &106 &     &  \\
\hline
\end{tabular}
Notes and units as in Table~\ref{tab:props-cluster}.
\end{table*}

\end{document}